\documentclass[12pt, draftclsnofoot, onecolumn]{IEEEtran}

\usepackage{graphicx}
\usepackage{amsfonts}
\usepackage{chngcntr}
\usepackage{times}
\usepackage{latexsym}
\usepackage{amssymb}
\usepackage{amsmath}
\usepackage{cite}
\usepackage{verbatim}
\usepackage{subfigure}
\usepackage{multirow}
\usepackage{lastpage}
\usepackage{calc}
\usepackage{eso-pic}
\usepackage{tabularx}
\usepackage{ifthen}
\usepackage{epstopdf}

\newtheorem{theorem}{Theorem}
\newcounter{as}
\newcounter{le}
\newcounter{cor}[theorem]
\newcounter{re}

\newcounter{pro}
\newcounter{law}

\newtheorem{scalinglaw}[law]{Scaling law}

\newtheorem{assumption}[as]{Assumption}

\newtheorem{corollary}[cor]{Corollary}
\counterwithin{cor}{theorem}

\newtheorem{lemma}[le]{Lemma}

\newtheorem{proposition}[pro]{Proposition}

\newtheorem{remark}[re]{Remark}
\newenvironment{proof}{ \textbf{Proof:} }{ \hfill $\Box$}

\def\bb0{{\mathbb{0}}}

\def\bb{{\mathbf{b}}}

\def\b0{{\mathbf{0}}}

\def\sf0{{\mathsf{0}}}

\usepackage{acronym}
\usepackage{url}

\acrodef{UHF}{ultra high frequency}
\acrodef{mmWave}{millimeter wave}
\acrodef{LOS}{line-of-sight}
\acrodef{NLOS}{non-line-of-sight}
\acrodef{SINR}{signal-to-noise-and-interference ratio}
\acrodef{RMS}{root mean square}
\acrodef{IID}{identically and independently distributed}
\acrodef{MIMO}{multiple-input and multiple-output}
\acrodef{FBR}{front-to-back ratio}
\acrodef{LTE}{long-term evolution}
\acrodef{PPP}{Poisson point process}
\acrodef{TDD}{time-division duplex}
\acrodef{SNR}{signal-to-noise ratio}
\acrodef{SIR}{signal-to-interference ratio}
\acrodef{MRC}{maximum ratio combining}
\acrodef{ZF}{zero-forcing}
\acrodef{MMSE}{minimum mean square error}
\acrodef{CSI}{channel state information}
\allowdisplaybreaks
\begin{document}

\title{Analyzing Uplink SINR and Rate in Massive MIMO Systems Using Stochastic Geometry}

\author{
Tianyang Bai and Robert W. Heath, Jr.
\thanks{The authors are with The University of Texas at Austin, Austin, TX, USA.
(email: tybai@utexas.edu, rheath@utexas.edu) This work is supported by the National Science Foundation under Grant Nos. 1218338, 1319556, and 1514257.
Parts of the results on the analysis of MRC receivers was presented at the 2015 IEEE Global Conference on Communications \cite{Bai2015globel}. }}
\maketitle
\begin{abstract}
This paper proposes a stochastic geometry framework to analyze the SINR and rate performance in a large-scale uplink massive MIMO network. Based on the model, expressions are derived for spatial average SINR distributions over user and base station distributions with maximum ratio combining (MRC) and zero-forcing (ZF) receivers. We show that using massive MIMO, the uplink SINR in certain urban marco-cell scenarios is limited by interference. In the interference-limited regime, the results reveal that for MRC receivers, a super-linear (polynomial) scaling law between the number of base station antennas and scheduled users per cell preserves the uplink SIR distribution, while a linear scaling applies to ZF receivers. ZF receivers are shown to outperform MRC receivers in the SIR coverage, and the performance gap is quantified in terms of the difference in the number of antennas to achieve the same SIR distribution. Numerical results verify the analysis. It is found that the optimal compensation fraction in fractional power control to optimize rate is generally different for MRC and ZF receivers. Besides, simulations show that the scaling results derived from the proposed framework apply to the networks where base stations are distributed according to a hexagonal lattice.
\end{abstract}

\section{Introduction}
Massive \ac{MIMO} is an approach to increase the area spectrum efficiency in 5G cellular systems \cite{Marzetta2010,Larsson2014,Lu2014,Boccardi2014}. By deploying large-scale antenna arrays, base stations can use multi-user \ac{MIMO} to serve a large number of users and provide high cell throughput \cite{Marzetta2010,Larsson2014,Lu2014,Boccardi2014}. In this paper, we focus on the defacto massive \ac{MIMO} systems operated below 6 GHz, where pilot-aided channel estimation is performed in the uplink, and pilots are reused across cells to reduce the training overhead \cite{Marzetta2010,Larsson2014,Lu2014,Boccardi2014}. Prior work showed that when the number of base station antennas grows large, high throughput is achieved through simple signal processing, and that the asymptotic performance of massive \ac{MIMO} (in the limit of the number of base station antennas) is limited by pilot contamination \cite{Marzetta2010}.

In this paper, we derive the \ac{SIR} distribution for the uplink of a massive MIMO network with \ac{MRC} and \ac{ZF} receivers, for a random base station topology. The performance with \ac{MRC} and \ac{ZF} beamforming in terms of \ac{SINR}, spectrum efficiency, and energy efficiency was examined in a simple network topology, e.g. in \cite{Yang2013,Marzetta2010,Krishnan2014,Hoydis2013,Li2012,Ngo2013a}, where the SIR and rate expressions were conditioned on specific user locations or equivalently the received power for each user. The conclusions drawn from the conditional expression, however, need not apply to the spatial average system-level performance due to the difference in users' path losses. For example, the linear scaling between the number of users and antennas examined in \cite{Hoydis2013} does not maintain the uplink \ac{SIR} distribution, as will be shown in our analysis. This motivates the analysis of the spatial average performance over different base station and user distributions in large-scale massive \ac{MIMO} networks, which was mainly studied using Monte Carlo simulations in prior work \cite{Bjornson2014aa,Atzeni2015}.

Stochastic geometry provides a powerful tool to analyze system-level performance in a large-scale network with randomly distributed base stations and users. Assuming a single antenna at each base station, the spatial average downlink SIR and rate distributions were derived for a network with \ac{PPP} distributed base stations, and were shown a reasonable fit with simulations using real base station data \cite{Andrews2011b}. The stochastic geometry framework in \cite{Andrews2011b} was further extended to analyze the performance of \ac{MIMO} networks: the downlink SIR and rate of multi-user \ac{MIMO} cellular system were analyzed, e.g. in \cite{Dhillon2013,Chandrasekhar2009,Wu2015,Li2013} assuming perfect \ac{CSI}, and in \cite{Kountouris2012} with quantized \ac{CSI} from limited feedback. For uplink analysis, prior work \cite{Novlan2013,ElSawy2014,Singh2014} showed that the uplink and downlink \ac{SIR} follows different distributions, due to the difference in network topology. In \cite{Singh2014}, a stochastic geometry uplink model was proposed to take account the pairwise correlations in the user locations, where the \ac{SIR} distributions derived based on the analytical model were shown a good fit with the simulations. The prior results in \cite{Andrews2011b,Dhillon2013,Chandrasekhar2009,Kountouris2012,Wu2015,Li2013,Novlan2013,ElSawy2014,Singh2014}, however, do not directly apply to analyze uplink massive MIMO networks, as (i) they did not take account for the effects of pilot contamination, which becomes a limiting factor with large numbers of antennas \cite{Marzetta2010}; (ii) the analysis in \cite{Andrews2011b,Dhillon2013,Chandrasekhar2009,Kountouris2012,Wu2015,Li2013} was intended for downlink performance, which follows different distributions from the uplink network; and (iii) the results in \cite{Dhillon2013,Chandrasekhar2009,Wu2015,Li2013} were intended for \ac{MIMO} networks with a few antennas, where the computational complexity for the analytical expressions grows with the number of antennas, and hinders the direct application to the massive \ac{MIMO} scenarios.

Stochastic geometry was also applied to study the asymptotic \ac{SIR} and rate in a massive \ac{MIMO} networks in \cite{Madhusudhanan2013,Bai2014h}, where the asymptotic SIR is shown to be approached with impractically large number of antennas, e.g. $10^4$ antennas. Related work in \cite{Liang2015} applied stochastic geometry to study the uplink interference in a massive \ac{MIMO} network. A linear scaling between the numbers of base station antennas and scheduled users was found to maintain the mean interference, which need not preserve the SIR distribution.

In this paper, we propose a stochastic geometry framework to derive the uplink \ac{SINR} and rate distributions in a large-scale cellular network using multi-user \ac{MIMO}. To model the uplink topology, we propose an exclusion ball model based on prior work \cite{Singh2014}, which simplifies the computation. Channel estimation error due to pilot contamination is also considered in the system model. The proposed framework also incorporates the fractional power control by compensating for a fraction of the path loss as in long term evolution (LTE) systems \cite{Xiao2006}. Based on the framework, we derive analytical expressions for the uplink SINR distribution for both \ac{MRC} and \ac{ZF} receivers in the massive \ac{MIMO} regime. Unlike prior work analyzing asymptotic performance with infinity antennas \cite{Madhusudhanan2013,Bai2014h}, the SINR coverage is examined as a function of the number of base station antennas and scheduled users per cell.

We apply the \ac{SINR} results to investigate the interference-limited case, as numerical results show that the impact of noise becomes minor in the rban macro-cell scenario with certain typical system parameters. We derive scaling laws between the number of base station antennas and scheduled users per cell to maintain the same uplink \ac{SIR} distributions. Unlike the linear scaling law examined in prior work \cite{Liang2015,Hoydis2013}, we find that a super-linear scaling is generally required for \ac{MRC} receivers to maintain the uplink SIR distributions, due to the near-far effect from intra-cell interference. For \ac{ZF} receivers, we show that a linear scaling law still holds, as the intra-cell interference is negligible. We use the scaling law results to quantify the performance gap between \ac{ZF} and \ac{MRC} receivers, in terms of the difference in the number of antennas to provide the same SIR distribution. The results show that \ac{ZF} receivers provides better \ac{SIR} coverage than \ac{MRC} receivers; the performance gap increases with the number of scheduled users in a cell, and is reduced with the fractional power control, as it mitigates the near-far effect from intra-cell interference. Simulations verify our analysis, and indicate that the scaling laws derived from the stochastic geometry framework also apply to the hexagonal model. Numerical results on rate also show that the average per user rate can be maximized by adjusting the compensation fraction, where the optimal fraction is around 0.5 for \ac{MRC} receivers, and 0.2 for \ac{ZF} receivers.

Our prior work in \cite{Bai2015globel} focused on the performance of \ac{MRC} receivers, and provided an expression for the \ac{SIR} distribution assuming no power control. In this paper, we incorporate thermal noise in the analysis, extend the results to the case of general fractional power control for MRC receivers, and analyze the performance of \ac{ZF} receivers.

This paper is organized as follows. We present the system model for network topology and channel assumptions in Section \ref{sec:system}. We analyze the performance of \ac{MRC} receivers in Section \ref{sec:MRC}, and that of \ac{ZF} receivers in Section \ref{sec:ZF}, followed by a performance comparison of two receivers in Section \ref{sec:comparison}. We present numerical results to verify the analysis in Section \ref{sec:numerical}, and conclude the paper in Section \ref{sec:conclusion}.

{\bf Notation:} We use the following notation throughout this paper: Bold lower-case letters $\mathbf{x}$ are used to denote vectors, and bold
upper-case letters $\mathbf{X}$ are used to denote matrices. We use $\mathbf{X}[:,k]$ to denote the k-th row of matrix $\mathbf{X}$, $\mathbf{X}^{*}$ as the Hermitian transpose of $\mathbf{X}$, and $\mathbf{X}^{\dagger}$ as the pseudo-inverse of $\mathbf{X}$. We use $\mathbb{E}$ to denote expectation, and $\mathbb{P}$ to denote probability.

\section{System Model}\label{sec:system}

In this section, we introduce the system model for an uplink massive MIMO cellular network. We focus on the networks operated in the sub-6 GHz band; the proposed model can be extended for massive MIMO at \ac{mmWave} frequencies by incorporating key differences in propagation and hardware constraints \cite{Bai2015}. Each base station is assumed to have $M$ antennas. In each time-frequency resource block, a base station can simultaneously schedule $K$ users in its cell. Let $X_\ell$ be the location of the $\ell$-th base station, $Y_\ell^{(k)}$ be the location of the $k$-th scheduled user in the cell of $\ell$-th base station, and $\mathbf h_{\ell\ell'}^{(k)}$ the channel vector from $X_\ell$ to $Y_{\ell'}^{(k)}$.

We consider a cellular network with perfect synchronization, and assume the following pilot-aided channel estimation in the uplink. In the uplink channel training stage, the scheduled users $Y_{\ell}^{(k)}$ send their assigned pilots $\mathbf t_k$, and base stations $X_\ell$ estimate the channels by correlating the corresponding pilots and using an \ac{MMSE} estimator; in the uplink data transmission, the base stations will apply either \ac{MRC} or \ac{ZF} receivers, based on the channel estimates derived from uplink pilots. Further, we assume the pilots $\{\mathbf{t}_k\}_{1\le k\le K}$ are orthogonal and fully reused in the network. Note that the system model assumption applies to general uplink multi-user \ac{MIMO} networks with pilot-aided channel estimation in the uplink, including but not limited to the \ac{TDD} massive MIMO \cite{Marzetta2010}.

Now, we introduce the channel model assumptions. The channel is assumed to be constant during one resource block and fades independently from block to block. Moreover, we apply a narrowband channel model, as frequency selectivity in fading can be minimized by techniques like orthogonal frequency-division multiplexing (OFDM) and frequency domain equalization \cite{Goldsmith2005}. We express the channel vector $\mathbf{h}^{(k)}_{\ell n}\in\mathcal{C}^{M\times1}$ as
\begin{align}\label{eqn:UHFsmallscale}
\mathbf{h}^{(k)}_{\ell n}=\left(\beta^{(k)}_{\ell n}\right)^{1/2}\mathbf \Phi_{\ell n}^{(k)1/2}\mathbf{w}^{(k)}_{\ell n},
\end{align}
where $\beta^{(k)}_{\ell n}$ is the large-scale path loss, $\mathbf{w}^{(k)}_{\ell n}\in\mathcal{C}^{M\times1}$ is a Gaussian vector with the distribution $\mathcal{CN}(\mathbf{0},\mathbf{I}_M)$ for Rayleigh fading, and $\mathbf{\Phi}^{(k)}_{\ell n}\in\mathcal{C}^{M\times M}$ is the covariance matrix to account for potential correlations in small-scale fading. In this paper, we focus on the case of \ac{IID} channel fading channels, i.e., $\Phi_{\ell n}^{(k)}=\mathbf{I}_M$. The incorporation of spatial correlations in fading is deferred to future work.

The large-scale path loss gain $\beta_{\ell n}^{(k)}$ is computed as
\begin{align}
\beta_{\ell n}^{(k)}=C\left(R_{\ell n}^{(k)}\right)^{-\alpha},
\end{align}
where $C$ is a constant determined by the carrier frequency and reference distance, $\alpha>2$ is the path loss exponent.


Next, we introduce the network topology assumptions based on stochastic geometry. We assume the base stations are distributed as a \ac{PPP} with a density $\lambda_\mathrm{b}$. A user is assumed to be associated with the base station that provides the minimum path loss signal. In this paper, each base station is assumed to serve $K$ scheduled users that are independently and uniformly distributed in its Voronoi cell \cite{Bjornson2015ab}. The assumption is equivalent to that in prior work \cite{Singh2014}, where the scheduled user process is obtained by (i) generating an overall user process as a PPP, and (ii) randomly selecting $K$ out of associated users in each cell as its scheduled users, under the full buffer assumption that the overall user process is sufficiently dense, such that each base station has at least $K$ candidate users in the cell. Without loss of generality, a typical {\it scheduled user} $Y_0^{(1)}$ is fixed at the origin, and its serving base station $X_0$ is denoted as the {\it tagged base station} in this paper. We will investigate the SINR and rate performance at this typical user.

Now we focus on modeling the distribution of {\it scheduled user process} in a resource block. For $1\le k\le K$, the $k$-th scheduled user $Y_{\ell}^{(k)}$ in each cell is assigned with the same pilot $\mathbf t_k$. Let $\mathcal{N}_\mathrm{u}^{(k)}$ be the point process formed by the locations of the $k$-th scheduled users $Y_{\ell}^{(k)}$ from each cell. Note that the scheduled user process $\mathcal{N}_\mathrm{u}^{(k)}$ is non-stationary (also non-PPP), as their locations are correlated with the base station process, and the presence of one scheduled user using $\mathbf t_k$ prohibits the others' in the same cell \cite{Novlan2013,Singh2014,ElSawy2014}. Unfortunately, the correlations in the scheduled users' locations make the exact analysis intractable. In \cite{Singh2014}, the authors proposed an uplink model to account for the pairwise correlations, where the other-cell scheduled users for base station $X_0$ in $\mathcal{N}_\mathrm{u}^{(k)}$ is modelled as an inhomogeneous PPP with a density function of
\begin{align}\label{eqn:lambda}
\lambda_\mathrm{u}(r)=\lambda_\mathrm{b}\left(1-\mathrm{e}^{-\lambda_\mathrm{b}r^2}\right),
\end{align}
where $r$ is the distance to base station $X_0$. To further simplify the analysis, e.g., the computation in (\ref{eqn:approx_new1}) and (\ref{eqn:approx_new2}), we propose an {\it exclusion ball approximation}, as a first-order approximation of the model in \cite{Singh2014}, on the distribution of the scheduled user process $\mathcal{N}_\mathrm{u}^{(k)}$ as follows.

\begin{assumption}\label{approx:1}The following assumptions are made to approximate the exact scheduled users' process $\mathcal{N}_\mathrm{u}^{(k)}$.
\begin{enumerate}
\item The distances $R_{\ell\ell}^{(k)}$ from a user to their associated base stations are assumed to be \ac{IID} Rayleigh random variables with mean $0.5\sqrt{1/\lambda_\mathrm{b}}$ \cite{Novlan2013}.
\item The other-cell scheduled user process $\mathcal{N}_\mathrm{u}^{(k)}$ is modeled by a homogenous \ac{PPP} of density $\lambda_\mathrm{b}$ outside an exclusion ball centered at the tagged base station $X_0$ with a radius $R_\mathrm{e}$.
\item The scheduled users processes using different pilots $\mathcal{N}_\mathrm{u}^{(k)}$ and $\mathcal{N}_\mathrm{u}^{(k')}$ are assumed to be independent for $k\ne k'$.
\end{enumerate}
\end{assumption}
Note that in the exclusion ball model,  we equivalently use a step function $\lambda_\mathrm{b}\left(1-\mathbb{I}(r<R_\mathrm{e})\right)$ to approximate the density function in (\ref{eqn:lambda}), where $\mathbb{I}(\cdot)$ is the indicator function.

In this paper, we let $R_\mathrm{e}=\sqrt{1/\left(\pi\lambda_\mathrm{b}\right)}$ by matching the average number of the excluded points from a homogenous PPP of density $\lambda_\mathrm{b}$ in the step function and in (\ref{eqn:lambda}), i.e., by letting $\lambda_\mathrm{b}\pi R_\mathrm{e}^2=2\pi\lambda_\mathrm{b}\int_{0}^{\infty}\mathrm{e}^{-\lambda_\mathrm{b}\pi r^2}r\mathrm{d}r=1$. An alternative explanation for our choice of $R_\mathrm{e}$ is to let the size of the exclusion ball $\pi R_\mathrm{e}^2$ equal the average cell size $1/\lambda_\mathrm{b}$ \cite{Baccelli2009a}. In Section \ref{sec:numerical}, we show that the SINR distributions derived based on the exclusion ball assumption, as well as the approximations made in our subsequent derivation, match well with the simulation using the exact user distribution.

%

Fractional power control, as used in the LTE systems \cite{Xiao2006}, is assumed in both the uplink training and uplink data stages: the user $Y_\ell^{(k)}$ transmits with power
\begin{align}\label{eqn:power}
P^{(k)}_{\ell}=P_\mathrm{t}\left(\beta^{(k)}_{\ell\ell}\right)^{-\epsilon},
\end{align}
where $\beta^{(k)}_{\ell\ell}$ is the path loss in the corresponding signal link, $\epsilon\in[0,1]$ is the fraction of the path loss compensation, and $P_\mathrm{t}$ is the open loop transmit power with no power control. We omit the constraint on the maximum uplink transmit power for simplicity; the constraint can be incorporated by applying the truncated channel inversion power control model \cite{ElSawy2014} to determine the transmit power. We note that ignoring the maximum transmit power constraint increases the average transmit power, and reduces the impact of noise. The incorporation of more complicated power control algorithms is deferred to future work. The noise power is denoted as $\sigma^2$.

In the uplink training stage, after correlating the received training signal with the corresponding pilot, base station $X_0$ has an observation of the channel $\mathbf{h}_{00}^{(1)}$ as
\begin{align*}
\mathbf{u}_{00}^{(1)}=\sqrt{P^{(1)}_{0}}\mathbf{h}_{00}^{(1)}+\sum_{\ell>0}\sqrt{P^{(1)}_{\ell}}\mathbf{h}_{\ell0}^{(1)}+\mathbf{n}_\mathrm{t},
\end{align*}
where $\mathbf{n}_\mathrm{t}$ is the noise vector in the training stage following the distribution $\mathcal{CN}\left(\mathbf{0},\frac{\sigma^2}{K}\mathbf{I}_M\right)$.

We assume for $\ell>0$, the large-scale path losses $\beta^{(1)}_{0\ell}$ are perfectly known to base station $X_0$. Since the channels are assumed to be \ac{IID} Rayleigh fading, the channel $\mathbf{h}_{00}^{(1)}$ is estimated by an \ac{MMSE} estimator as
\begin{align}\label{eqn:MMSE1}
\bar{\mathbf{h}}_{00}^{(1)}=\frac{\sqrt{P^{(1)}_{0}}\beta^{(1)}_{00}}{\sum_{\ell}P^{(1)}_{\ell}\beta^{(1)}_{0\ell}+\frac{\sigma^2}{K}}\mathbf{u}_{00}^{(1)},
\end{align}
where $\bar{\mathbf{h}}_{00}^{(1)}$ is the estimation of $\mathbf{h}_{00}^{(1)}$. Due to the orthogonality principle, the channel vector $\mathbf{h}_{00}^{(1)}$ can be decomposed as
\begin{align}
\mathbf{h}_{00}^{(1)}=\bar{\mathbf{h}}_{00}^{(1)}+\hat{\mathbf{h}}_{00}^{(1)},
\end{align}
where $\hat{\mathbf{h}}_{00}^{(1)}$ is the estimation error following the distribution $\mathcal{CN}\left(\mathbf{0},\beta_{00}^{(1)}\left(1-\frac{P^{(1)}_{0}\beta^{(1)}_{00}}{\sum_{\ell}P^{(1)}_{\ell}\beta^{(1)}_{0\ell}+\frac{\sigma^2}{K}}\right)\mathbf{I}\right)$.

Let $s^{(k)}_\ell$ be the uplink data symbol for user $Y_\ell^{(k)}$ with $\mathbb{E}\left[|s^{(k)}_\ell|^2\right]=P^{(k)}_\ell$. In uplink data transmission, base station $X_\ell$ is assumed to use the combiner vector $\mathbf{g}_{\ell \ell}^{(k)}$ to decode $s^{(k)}_\ell$ from $Y_\ell^{(k)}$, based on the channel estimate $\bar{\mathbf{h}}_{\ell \ell}^{(k)}$. Then, at base station $X_0$, the decoded symbol $\hat{s}^{(1)}_0$ for the typical user $X_{0}^{(1)}$ is
\begin{align}\label{eqn:symbol}
\hat{s}^{(1)}_0=\mathbf{g}_{00}^{(1)*}\bar{\mathbf{h}}_{00}^{(1)}s^{(k)}_\ell+\underbrace{\mathbf{g}_{00}^{(1)*}\hat{\mathbf{h}}_{00}^{(1)}s^{(k)}_\ell+\sum_{(\ell,k)\ne(0,1)}\mathbf{g}_{00}^{(1)*}\mathbf h_{0\ell}^{(k)}s_{\ell}^{(k)}+\mathbf{g}_{00}^{(1)*}\mathbf{n}_\mathrm{u}}_{\mbox{unknown at base station}},
\end{align}
where $\mathbf{n}_\mathrm{u}\in\mathcal{C}^{M\times 1}$ is the thermal noise vector in the uplink data transmission.
Treating the unknown terms at base station $X_0$ as uncorrelated additive noise, the uplink \ac{SINR} for the typical user $Y_{0}^{(1)}$ is
\begin{align}\label{eqn:UHF_UL_SIR}
\mathrm{SINR}=\frac{P_{0}^{(1)}| \mathbf{g}_{00}^{(1)*}\bar{\mathbf{h}}_{00}^{(1)}\mathbf |^2}{P_{0}^{(1)}\mathbb{E}| \mathbf{g}_{00}^{(1)*}\hat{\mathbf{h}}_{00}^{(1)} |^2+\sum_{(\ell,k)\ne(0,1)}P_{\ell}^{(k)}\mathbb{E}|\mathbf{g}_{00}^{(1)*}\mathbf h_{0\ell}^{(k)}|^2+|\mathbf{g}_{00}^{(1)}|^2\sigma^2},
\end{align}
where the expectation operator is taken over the channel estimation error and small-scale fading in the interference links. We will investigate the SINR distributions for \ac{MRC} and \ac{ZF} receivers in the following sections.

The proposed system model represents a simple multi-user \ac{MIMO} systems in which the \ac{SINR} expression can be analyzed using stochastic geometry. In the following sections, we will study the uplink \ac{SINR} and rate distributions for \ac{MRC} and \ac{ZF} receivers, when the number of base station antennas is large.

\section{Performance Analysis for \ac{MRC} Receivers}\label{sec:MRC}
In this section, we derive an approximate \ac{SINR} distribution in an uplink multi-user \ac{MIMO} network, where the approximation becomes tight in the massive \ac{MIMO} regime, e.g. when $M>64$. Then, we focus on the interference limited case, as numerical results show that the uplink SINR is dominated by the interference in certain urban macro-cell scenarios. We derive a scaling law between the number of users and antennas that maintains the uplink SIR distribution at the typical user. Finally, we present a method to compute the per-user achievable rate and cell throughput, based on the SINR distribution.
\subsection{SIR Coverage Analysis}\label{MRC_SIR}
Now we investigate the uplink SINR coverage based on the system model. With \ac{MRC} receivers, we assume that base station $X_0$ applies the combining vector $\mathbf{g}_{00}^{(k)}$ as a scaled version of the channel estimate $\bar{\mathbf{h}}_{00}^{(1)}$ to decode the signal from $Y_{00}^{(1)}$:
\begin{align}\label{eqn:mrc_bf}
\mathbf{g}_{00}^{(1)}=\frac{\sum_{\ell}P^{(1)}_{\ell}\beta^{(1)}_{0\ell}+\frac{\sigma^2}{K}}{\sqrt{P^{(1)}_{0}}\beta^{(1)}_{00}}\bar{\mathbf{h}}_{00}^{(1)}=\mathbf{u}_{00}^{(1)}.
\end{align}
Note the scaling on the combining vector is intended to simplify expressions, and will not change the \ac{SINR} distribution. Then, using the combining vector in (\ref{eqn:mrc_bf}), the \ac{SINR} expression can be simplified in (\ref{eqn:SIR_MRC_simplified}) as
{\small\begin{align}\label{eqn:SIR_MRC_simplified}
\mathrm{SINR}=\frac{(M+1)\left(\beta_{00}^{(1)}\right)^{2(1-\epsilon)}}{M\Delta_2^{(1)}+\left(\beta_{00}^{(1)}\right)^{1-\epsilon}\Delta_1^{(1)}+\left(\sum_{k=2}^{K}\left(\beta_{00}^{(k)}\right)^{1-\epsilon}+\sum_{k=1}^{K}\Delta_1^{(k)}\right)\left(\left(\beta_{00}^{(1)}\right)^{1-\epsilon}+\Delta_1^{(1)}\right)}
\end{align}
}where
$\Delta_1^{(k)}=\sum_{\ell>0}\left(\beta^{(k)}_{\ell\ell}\right)^{-\epsilon}\beta^{(k)}_{0\ell}+\frac{\sigma^2}{KP_\mathrm{t}},$ and
$\Delta_2^{(k)}=\sum_{\ell>0}\left(\beta^{(k)}_{\ell\ell}\right)^{-2\epsilon}\left(\beta^{(k)}_{0\ell}\right)^2.
$
The derivation to obtain (\ref{eqn:SIR_MRC_simplified}) is given in Appendix A. Note that $\Delta_1^{(k)}$ and $\Delta_2^{(k)}$ correspond to the sum of certain interference terms from other-cell users.

Next, we denote the exact \ac{SINR} distribution for (\ref{eqn:SIR_MRC_simplified}) (using the exact scheduled user distribution defined in Section \ref{sec:system} but not the exclusion ball assumption) as $\mathbb{P}(\mathrm{SINR}>T)$.  Due to pilot contamination, the combining vector $\mathbf{g}_{00}^{(k)}$ is correlated with certain interference channel vectors as shown in (\ref{eqn:mrc_bf}). As a result, the denominator in (\ref{eqn:SIR_MRC_simplified}) contains cross-products of the path losses from different interferers. Moreover, different cross-product terms in the denominator of (\ref{eqn:SIR_MRC_simplified}) can be correlated, as they may contains common path loss terms, which renders the exact derivation of $\mathbb{P}(\mathrm{SINR}>T)$ intractable. Therefore, we compute an approximate \ac{SINR} distribution $\bar{\mathbb{P}}(\mathrm{SINR}>T)$, which we argue in Section \ref{sec:numerical} is a good match for $\mathbb{P}(\mathrm{SINR}>T)$, in Theorem \ref{thm:MRC_SIR}.

\begin{theorem}[\ac{MRC} SINR]\label{thm:MRC_SIR} In the proposed massive MIMO networks, an approximate uplink SINR distribution with \ac{MRC} receivers can be computed as
\begin{align}\label{eqn:MRC_distribution}
\bar{\mathbb{P}}(\mathrm{SINR}>T)=
\sum_{n=1}^{N}{{N}\choose{n}}(-1)^{n+1}\int_{0}^{\infty}\mathrm{e}^{-t- \ell T\eta C_1t^{\alpha(1-\epsilon)}-\ell T\eta C_2t^{\frac{\alpha}{2}(1-\epsilon)}}C_3(t)\mathrm{d}t,
\end{align}
where $N$ is the number of terms used in the calculation, $\eta=N(N!)^{-\frac{1}{N}}$, $C_{\sigma^2}=\frac{\sigma^2}{KP_\mathrm{t}C^{1-\epsilon}\left(\lambda_\mathrm{b}\pi\right)^{\frac{\alpha(1-\epsilon)}{2}}}$, $C_1=\frac{K+1}{M+1}\left(\frac{2\Gamma^\alpha(\frac{\epsilon}{2}+1)}{(\alpha-2)}+C_{\sigma^2}\right)$, $C_2=\frac{M\Gamma^\alpha(\epsilon + 1)}{(M+1)(\alpha - 1)}+\frac{K}{M+1}\left(\frac{2\Gamma^{\alpha}(\frac{\epsilon}{2}+1)}{(\alpha-2)}+C_{\sigma^2}\right)^2$,
\begin{align*}
C_3(t)&=\left(\int_{0}^{\infty}\mathrm{e}^{-u-u^{-\frac{\alpha}{2}(1-\epsilon)}\ell T\eta C_4(t)}\mathrm{d}u\right)^{K-1}\\
&\approx\left(1-\ell \eta TC_4(t)\int_{0}^{\infty}\frac{\mathrm{e}^{-u}}{\ell T\eta C_4(t)+u^{-\frac{\alpha}{2}(1-\epsilon)}}\mathrm{d}u\right)^{K-1},
\end{align*}
$C_4(t)=\frac{1}{M+1}\left(\left(\frac{2\Gamma^\alpha(\frac{\epsilon}{2} + 1)}{\alpha-2}+C_{\sigma^2}\right)t^{\alpha(1-\epsilon)}+t^{\frac{\alpha}{2}(1 -\epsilon)}\right)$, and $\Gamma(\alpha)=\int_{0}^{\infty}\mathrm{e}^{-t}t^{\alpha-1}\mathrm{d}t$ is the gamma function.
\end{theorem}
\proof
See Appendix B.
\endIEEEproof

Besides the exclusion ball approximation, the main approximation in Theorem \ref{thm:MRC_SIR} is to replace certain out-of-cell interference terms by their means in (\ref{eqn:approx_new1}) and (\ref{eqn:approx_new2}). The approximation results in a minor error in the SINR distribution, as (i) with $K$ users in a cell, the intra-cell interference dominates the out-of-cell interference with high probability; (ii) with large antenna arrays, the ratio of the signal power to certain out-of-cell interference power terms, e.g. the terms in $\Delta_1^{(k)}$, decays as $\frac{1}{M}$. In Section \ref{sec:numerical}, using $N\ge5$ terms, the distribution $\bar{\mathbb{P}}(\mathrm{SINR}>T)$ computed in Theorem \ref{thm:MRC_SIR} is shown to be a good match with the SINR distribution ${\mathbb{P}}(\mathrm{SINR}>T)$ from Monte Carlo simulations. In addition, the error of the approximation becomes more prominent with a smaller noise power, as all the approximations are made with respect to the interference distribution. The expression is intended for the massive MIMO regime when $M\gg 1$, as the error of the approximations decays with $\frac{1}{M}$. In simulations, we find that the results in the theorem generally applies to the multi-user MIMO networks with not-so-large $M$, e.g. the case of $(M,K)=(10,2)$.

 In Theorem \ref{thm:MRC_SIR}, the noise power is taken account by the parameter $C_{\sigma^2}=\frac{\sigma^2}{KP_\mathrm{t}C^{1-\epsilon}\left(\lambda_\mathrm{b}\pi\right)^{\frac{\alpha(1-\epsilon)}{2}}}$, which shows that the impact of noise on the SINR is reduced with a larger number of scheduled users per cell $K$, a higher base station density $\lambda_\mathrm{b}$, a smaller path loss $\alpha$, and a larger power control parameter $\epsilon$. Besides, the impact of noise goes down with larger $M$, as in the expressions for $C_1$ and $C_2$, the noise parameter $C_{\sigma^2}$ is divided by $(M+1)$.

Next, we focus on the performance of interference-limited networks. We will show in Section \ref{sec:numerical} that the impact of noise is negligible in certain urban macro-cell cases with $M=64$ antennas at base stations. Then, the general expression in Theorem \ref{thm:MRC_SIR} can be further simplified in the following special cases.

{\bf Case 1 (Full power control, $\epsilon=1$):} In this case, the transmitting power at scheduled user $Y_\ell^{(k)}$ is adjusted to compensate for the full path loss, i.e., $P_\ell^{(k)}=P_\mathrm{t}\beta_{\ell\ell}^{(k)}$, such that a base station receives equal signal powers from all of its associated users. When $\epsilon=1$, the SIR distribution can be simplified as in the following corollary.
\begin{corollary}\label{cor:fullpower}
With $\epsilon=1$ and $\sigma^2=0$, the approximate SIR distribution can be computed as
\begin{align}\label{eqn:MRC_distribution_full}
\bar{\mathbb{P}}(\mathrm{SIR}>T)=
\sum_{n=1}^{N}{{N}\choose{n}}(-1)^{n+1}\mathrm{e}^{- T\eta\ell\left(\frac{C_5K+\Gamma^\alpha(1.5)}{M+1}+\frac{1}{\alpha-1}\right)},
\end{align}
where $C_5=\frac{4\Gamma^{2\alpha}(1.5)+(\alpha^2-4)\Gamma^\alpha(1.5)}{(\alpha-2)^2}$.
\end{corollary}

Based on Corollary \ref{cor:fullpower}, a linear scaling law between the number of users and antennas is observed as follows.

\begin{corollary}\label{cor:full_scaling}
With $\epsilon=1$ and $\sigma^2=0$, to maintain the uplink SIR distribution unchanged, the scaling law between the number of base station antennas $M$ and users per cell $K$ is approximately
\begin{align}
(M+1)\sim\left(K+\frac{\Gamma^{\alpha}(1.5)}{C_5}\right)\approx K.
\end{align}
\end{corollary}
Note that when $\epsilon=1$, the linear scaling law matches prior results in \cite[Sec. IV]{Hoydis2013}, where the path loss to all associated users in the typical cell was assumed to be identical. The linear scaling law, however, does not apply to other cases with $\epsilon<1$, e.g. in the following case without power control.

{\bf Case 2 (No power control, $\epsilon=0$):} In this case, the fraction of the path loss compensation is $\epsilon=0$. Then, the uplink SIR can be evaluated as follows.
\begin{theorem}\label{thm:finite}
With $\epsilon=0$ and $\sigma^2=0$, an approximate uplink SIR distribution can be calculated as
\begin{align}\label{eqn:uplink_IID}
\bar{\mathbb{P}}(\mathrm{SIR}>T)=
\sum_{n=1}^{N}{{N}\choose{n}}(-1)^{n+1}\int_{0}^{\infty}\mathrm{e}^{-(\mu\Gamma(1-2/\alpha)(n\eta T)^{2/\alpha}+1) t-\frac{n\eta T}{\alpha-1}t^\alpha}\mathrm{d}t,
\end{align}
where $N$ is the number of terms used in the computation, and $\mu=\frac{K}{(M+1)^{2/\alpha}}$.
\end{theorem}
\proof
The proof is similar to that in \cite[Appendix A]{Bai2015globel}.
\endIEEEproof

We will show in Section \ref{sec:numerical} that Theorem \ref{thm:finite} provides a tight approximation of the exact SIR distribution $\mathbb{P}(\mathrm{SIR}>T)$, when $N\ge5$ terms are used. Moreover, note that in (\ref{eqn:uplink_IID}), the number of antennas $M$ and the number of scheduled users per cell $K$ only affect the value of $\mu$. Therefore, by Theorem \ref{thm:finite}, in the no power control case, we observe the following scaling law to maintain SIR.
\begin{corollary}
Assuming no power control, the approximate scaling law to maintain the same uplink SIR distribution is $(M+1)\sim K^{\alpha/2}$, which is a superlinear polynomial scaling when $\alpha>2$.
\end{corollary}

In the case of no power control, the difference in the path losses between the typical user and the intra-cell interferers affect the SIR distribution, and thus the scaling law to maintain the SIR becomes a function of the path loss exponent. The super-linearity in the scaling law can be explained by the near-far effect of the intra-cell interference from multiple users in a cell. With no power control, the cell edge users will receive weaker signals than the cell center user. With a uniform user distribution in a cell, the typical user will be more likely to be located at the cell edge. When increasing the number of scheduled users $K$ in a cell, the probability that the interference from a cell-center interferer dominates the signal from the typical user increases. Therefore, compared with the linear scaling law with full power control ($\epsilon=1$) where such near-far effect is mitigated, more antennas will be needed in the no power control case to reduce the intra-cell interference, and preserve the SIR distribution, when increasing $K$.

Next, we focus on the scaling law in the general fractional power control case with $\epsilon\in(0,1)$. It is difficult to derive the exact scaling law directly from the expression (\ref{eqn:MRC_distribution}), due to the integral form. Since with the fractional power control, the equivalent path loss exponent in the signal link linearly scales with $\epsilon$, we propose the following approximate scaling law by linearly fitting the exponent $s$ of the scaling law $(M+1)\sim K^s$, based on two special cases of $\epsilon$: by Corollary \ref{cor:full_scaling}, when $\epsilon=1$, $s=1$; and by Theorem \ref{thm:finite}, when $\epsilon=0$, $s=\frac{\alpha}{2}$. Therefore, for general $0<\epsilon<1$, the linearly fitted exponent of the scaling law $s$ is given as follows.

\begin{scalinglaw}\label{cor:conj}
With fractional power control, the scaling law between $M$ and $K$ is approximately $(M+1)\sim K^s$, where the exponent of the scaling law is $s=\frac{\alpha}{2}(1-\epsilon)+\epsilon$.
\end{scalinglaw}

Scaling law \ref{cor:conj} reveals that a (superlinear) polynomial scaling law between $K$ and $M$ is required to maintain uplink SIR distribution, for a general $\epsilon<1$. The results in Scaling law \ref{cor:conj} are verified  by numerical simulations in Section \ref{sec:numerical}.
\subsection{Rate Analysis}\label{sec:rate_MRC}
In this section, we apply the SINR results to compute the achievable rate. First, we define the average achievable spectrum efficiency at a typical user as
\begin{align}
\tau_0=\mathbb{E}\left[\log_2\left(1+{\min\{\mathrm{SINR},T_{\max}\}}\right)\right],
\end{align}
where $T_{\max}$ is a SINR distortion threshold determined by limiting factors like distortion in the radio frequency front-end. By \cite[Section III-C]{Bai2014}, given the SINR distribution $\mathbb{P}(\mathrm{SINR}>T)$, the average achievable spectrum efficiency can be computed as
$
\tau_0=\frac{1}{\ln(2)}\int_{0}^{T_{\max}}\frac{\mathbb{P}(\mathrm{SINR}>x)}{1+x}\mathrm{d}x.
$
To take account for the overhead, let $\psi$ be the fraction time for overhead. In this paper, for simplicity, we only consider the overhead due to uplink channel training, and compute the overhead fraction $\psi$ as $
\psi=\frac{T_\mathrm{t}}{T_\mathrm{c}}=\frac{K}{T_c},
$
where $T_\mathrm{t}$ and $T_\mathrm{c}$ are the length of channel training period and coherent time, in terms of the number of symbol time. The length of channel training is assumed to be equal to the number of scheduled users in a cell, as we assumed full reuse of orthogonal pilots throughout the network. Then the average achievable rate with the overhead penalty $\bar{\tau}_0$ equals
\begin{align}
\bar{\tau}_0=K(1-\psi)\tau_0,
\end{align}
Note that when ignoring thermal noise, the scaling law to maintain SINR distribution also maintains the average achievable rate $\tau_0$. When taking account for the training overhead penalty $\psi$, however, the scaling law will not keep $\bar{\tau}_0$ unchanged, as $1-\psi$ linearly decreases with $K$, unless $\psi$ is negligible, e.g. when the coherence time $T_\mathrm{c}\gg K$.
Next, we define the average cell throughput $\tau_\mathrm{cell}$, in terms of spectrum efficiency, as
\begin{align}
\tau_\mathrm{cell}=K(1-\psi)\tau_0.
\end{align}
We will examine the average cell throughput as a function of $M$ and $K$ in Section \ref{sec:numerical}. Before that, we continue to present the results for \ac{ZF} receivers in the next section.

\section{Performance Analysis with \ac{ZF} Receivers}

In this section, we will investigate the performance of \ac{ZF} receivers in \ac{IID} fading channels. First, we derive the SINR and rate distributions with \ac{ZF} receivers. Then, we apply the analytical results to compare the performance of \ac{MRC} and \ac{ZF} receivers in an interference-limited network. In particular, we aim to answer the question: compared with \ac{MRC} receivers, how many antennas can be saved by applying \ac{ZF} receivers, while keeping the same uplink SIR distribution.

\subsection{SINR Analysis of \ac{ZF} Receivers}\label{sec:ZF}
Now we begin to investigate the performance of \ac{ZF} receivers in an uplink massive \ac{MIMO} network. For \ac{ZF} receivers, we still focus on the case of \ac{IID} fading. To cancel the intra-cell interference, base station $X_\ell$ will apply the combining vector $\mathbf{g}_{\ell\ell}^{(k)}$ for user $Y_\ell^{(k)}$ as
\begin{align}\label{eqn:combinig_zf}
\mathbf{g}_{\ell\ell}^{(k)}=\mathbf{H}^{\dagger}_{\ell}[:,k],
\end{align}
where $\mathbf{H}_{\ell}=\left[\mathbf{u}_{\ell\ell}^{(1)},\mathbf{u}_{\ell\ell}^{(2)},\dots,\mathbf{u}_{\ell\ell}^{(K)}\right]\in\mathcal{C}^{M\times K}$ is the matrix of all estimated channels to the associated users in cell $X_\ell$, and $\mathbf{u}_{\ell\ell}^{(k)}=\frac{\sum_{\ell'}P^{(1)}_{\ell'}\beta^{(1)}_{\ell\ell'}+\frac{\sigma^2}{K}}{\sqrt{P^{(k)}_{\ell}}\beta^{(k)}_{\ell\ell}}\bar{\mathbf{h}}_{\ell\ell}^{(k)}$ is a scaled version of the channel estimate. The scaling in the channel estimates will not change the uplink SINR distribution, as it will only cause certain scaling in the corresponding combining vector. Similar to the case of \ac{MRC} receivers, the exact uplink SINR distribution is difficult to derive, as due to pilot contamination, the combining vector is correlated with certain interference channel vectors. Therefore, applying the same approximations in (\ref{eqn:approx_new1}) and (\ref{eqn:approx_new2}), we derive an approximate distribution for the uplink SINR expression in (\ref{eqn:UHF_UL_SIR}) for the typical user $X_0^{(1)}$ in the following theorem.

\begin{theorem}\label{thm:ZF_SIR}
With $M\gg K$ and \ac{ZF} receivers, an approximate uplink SINR distribution for the typical user can be calculated by
\begin{align}\label{eqn:uplink_ZF}
\bar{\mathbb{P}}(\mathrm{SINR}>T)=
\sum_{n=1}^{N}{{N}\choose{n}}(-1)^{n+1}\int_{0}^{\infty}\mathrm{e}^{-n\eta T\left(C_6 t^{\frac{\alpha}{2}(1-\epsilon)}+C_7 t^{\alpha(1-\epsilon)}\right)-t}\mathrm{d}t,
\end{align}
where the constant $C_6=C_9\left(\frac{1}{M-K+1}+\frac{1}{M+1}+\frac{M(K-1)}{(M-K+1)^2}\right)+\frac{M(K-1)C_8}{(M-K+1)^2}$, $$C_7=\frac{M}{M+1}\frac{\Gamma^\alpha(\epsilon+1)}{\alpha-1}+\left(\frac{1}{M+1}+\frac{(K-1)M}{(M-K+1)^2}\right)C_9^2+\frac{(K-1)M}{(M-K+1)^2}C_8C_9,$$
$C_8=\frac{2\Gamma^\alpha(\frac{\epsilon}{2}+1)+(\alpha-2)C_{\sigma^2}}{(\alpha-2)(1+C_{\sigma^2})+2\Gamma^\alpha(\frac{\epsilon}{2}+1)}$, $C_9=\frac{2\Gamma^{\alpha}(\frac{\epsilon}{2}+1)}{\alpha-2}+C_{\sigma^2}$, $C_{\sigma^2}=\frac{\sigma^2}{KP_\mathrm{t}C^{1-\epsilon}\left(\lambda_\mathrm{b}\pi\right)^{\frac{\alpha(1-\epsilon)}{2}}}$, $N$ is the number of terms used in the computation, and $\eta=N(N!)^{-\frac{1}{N}}$.
\end{theorem}
\proof See Appendix C. \endIEEEproof
Note that when $K=1$, the SINR distribution in (\ref{eqn:uplink_ZF}) for \ac{ZF} receivers is the same as that for \ac{MRC} receivers in (\ref{eqn:MRC_distribution_full}). We will verify the tightness of the approximation ${\mathbb{P}}(\mathrm{SINR}>T)\approx\bar{\mathbb{P}}(\mathrm{SINR}>T)$ by numerical simulation in Section \ref{sec:numerical}. We have the following remark on the applicable regime for Theorem \ref{thm:ZF_SIR}.
\begin{remark}
We need the condition $M\gg K$ in the proof, as the error in the approximation in (\ref{eqn:ZF_approx}) decays as $\frac{1}{M-K+1}$. In numerical simulations, we find that the approximate SINR distribution in Theorem \ref{thm:ZF_SIR} shows a good match with the simulations when $\frac{M}{K}\ge 3$ with $M\ge 10$. The same comment applies to Scaling law \ref{lem:scaling_ZF} below.
\end{remark}

Next, we focus on the interference-limited case. Based on Theorem \ref{thm:ZF_SIR}, we can derive an approximate scaling law between $M$ and $K$ to  maintain the SIR distribution in the region of $M\gg K$ as follows.
\begin{scalinglaw}\label{lem:scaling_ZF}
With {ZF} receivers and $\sigma^2=0$, the uplink SIR distribution of the typical user remains approximately unchanged when the number of antennas $M$ linearly scales with the number of users per cell $K$ as $(M+1)\sim K$.
\end{scalinglaw}
\proof Note that when $\sigma^2=0$, $C_{\sigma}=0$. The dependence on $M$ and $K$ in (\ref{eqn:uplink_ZF}) only occurs in the constants $C_6$ and $C_7$. Therefore, it is sufficient to show that a linear scaling between $M$ and $K$ (approximately) maintains the values of $C_6$ and $C_7$. Note that when $M\to\infty$, and $M\gg K$, the following limits hold: $\frac{1}{M+1}\to 0$, $\frac{1}{M-K+1}\to 0$, and $\frac{M}{M+1}\to 1$. Therefore, it follows that when keeping $\frac{K}{M+1}=t$,
$
\lim_{M\to\infty}C_6=\frac{(C_8+C_9)t}{1-t}$, and
$\lim_{M\to\infty}C_7=\frac{\Gamma^\alpha(\epsilon+1)}{\alpha-1}+\frac{t}{1-t}\left(\frac{4\Gamma^{2\alpha}(\frac{\epsilon}{2}+1)}{(\alpha-2)^2}+\frac{2C_8\Gamma^{\alpha}(\frac{\epsilon}{2}+1)}{\alpha-2}\right)$, which are invariant when $(M+1)$ linearly scales with $K$.\endIEEEproof

Compared with \ac{MRC} receivers, the near-far effect for users in a cell becomes minor with ZF receivers, as the intra-cell interference is largely suppressed. Therefore, a linear scaling law applies for \ac{ZF} receivers even without power control. Based on the SINR coverage results, the achievable rate per user and sum throughput can be computed following the same line as in Section \ref{sec:rate_MRC}. In the next section, we will use the derived results to compare the SIR coverage performance between \ac{MRC} and \ac{ZF} receivers in an interference-limited network.
\subsection{Comparison of SIR Coverage Performance}\label{sec:comparison}
Now assuming the network is interference-limited, we compare the SIR coverage between \ac{ZF} and \ac{MRC} receivers. Prior work \cite{Hoydis2013} showed that \ac{ZF} and \ac{MRC} receivers have the same asymptotic performance, both which are limited by the pilot contamination. The analysis in \cite{Liang2015} showed that by suppressing intra-cell interference, which turns to be more dominant than the out-of-cell interference, the \ac{ZF} receivers suffers from less interference than \ac{MRC} receivers. In this section, we make a quantitative comparison by answering the following question: in \ac{IID} fading channels, how many base station antennas $M_\mathrm{ZF}$ is needed for \ac{ZF} receivers to provide the same uplink SIR distribution as \ac{MRC} receivers with $M_\mathrm{MRC}$ antennas?

Based on Scaling law \ref{cor:conj} and Scaling law \ref{lem:scaling_ZF}, we have the following proposition to determine $M_\mathrm{ZF}$ to match the SIR coverage with MRC receivers.
\begin{proposition}\label{thm:comparison}
 Assuming $M_\mathrm{ZF}\gg K$, \ac{ZF} receivers with $(M_\mathrm{ZF}+1)=\xi (M_\mathrm{MRC}+1)$ antennas approximately provide the same uplink SIR distribution as \ac{MRC} receivers with $M_\mathrm{MRC}$ antennas in a massive MIMO networks, where the scaling factor $\xi=K^{-(\frac{\alpha}{2}-1)(1-\epsilon)}$, and $K$ is the number of scheduled users in a cell.
\end{proposition}
\proof
For the ease of notation, let $\mathrm{ZF}(M,K)$ and $\mathrm{MRC}(M,K)$ denote the uplink SIR distributions with \ac{ZF} and \ac{MRC} receivers of $M$ antennas, when serving $K$ users in a cell. By Scaling law \ref{lem:scaling_ZF}, when $M_\mathrm{ZF}\gg K$, $\mathrm{ZF}(M_\mathrm{ZF},K)\approx\mathrm{ZF}(\frac{M_\mathrm{ZF}+1}{K},1)$. Next, note that when $K=1$, i.e., with a single scheduled user in a cell, \ac{MRC} and \ac{ZF} receivers provide the same SIR coverage. Thus, it follows that $\mathrm{ZF}(M_\mathrm{ZF},K)=\mathrm{ZF}(\frac{M_\mathrm{ZF}+1}{K},1)=\mathrm{MRC}(\frac{M_\mathrm{ZF}+1}{K},1)$. Last, by Scaling law \ref{cor:conj}, $\mathrm{ZF}(M_\mathrm{ZF},K)\approx\mathrm{MRC}(\frac{M_\mathrm{ZF}+1}{K},1)\approx\mathrm{MRC}((M_\mathrm{ZF}+1)K^{(\frac{\alpha}{2}-1)(1-\epsilon)}-1,K)$.
\endIEEEproof

The condition $M_\mathrm{ZF}\gg K$ in the proposition is required to ensure the applicability of Scaling law \ref{lem:scaling_ZF}. In numerical simulations, the result is found to be a good approximation with $\frac{M_\mathrm{ZF}}{K}>3$. Note that the exponent of the scaling factor $-(\frac{\alpha}{2}-1)(1-\epsilon)$ is non-positive, which indicates we need $M_\mathrm{MRC}\ge M_\mathrm{ZF}$ to provide the same SIR coverage. Further, the scaling factor $\xi$ increases with the number of the scheduled user $K$, which reveals that the performance gap between \ac{MRC} and \ac{ZF} receivers grows with $K$. When $K$ increases, the mitigation of the intra-cell interference from $(K-1)$ users by \ac{ZF} receivers becomes more prominent to improve SIR coverage. In addition, Proposition \ref{thm:comparison} also shows that in terms of the SIR distribution, the performance gap reduces with larger $\epsilon$ in the power control scheme, as the scaling factor $\xi$ is a decreasing function of $\epsilon$. Simulations show that with $\epsilon=1$, only a minor gap exists between the SIR coverage curves for \ac{ZF} and \ac{MRC} receivers.

Last, we note that Proposition \ref{thm:comparison}, which is drawn based on the SIR distribution, need not extend to a general SINR distribution that is not dominated by interference; prior work \cite{Yang2013} showed that when the noise is not negligible, MRC receivers would provide a comparable or even better SINR, compared with the ZF receivers. In the following section, we will present numerical results to validate our analytical results.
\section{Numerical Results}\label{sec:numerical}
In this section, we verify our analytical results with numerical simulations, which follow the procedure as: (1) generating the base station process as a \ac{PPP} of density $\lambda_\mathrm{b}$; (2) generating the overall user process as a PPP of density $\lambda_{\mathrm{u,o}}$, where we use $\lambda_{\mathrm{u,o}}=60\lambda_\mathrm{b}$, unless otherwise specified; (3) associating the points in the overall user process to base stations, based on the minimal path loss rule, and then randomly scheduling $K$ out of the associated users in each cell as their scheduled users; (4) picking the base station closest to the origin as the tagged base station $X_0$, and its first scheduled user $Y_{0}^{(1)}$ as the typical user; (5) generating channel vectors as \ac{IID} Gaussian vectors, and computing the SINR for the iteration; (6) repeating the step (1)-(5) for 10,000 iterations, and computing the empirical distribution of the SINR at $Y_{0}^{(1)}$. For the simulations using hexagonal grids, we follow the same procedure except that the base station process is generated as a 19-cell hexagonal grid, and the tagged base station is the center cell. In addition, we will use $N=5$ terms when evaluating the analytical expressions.
\begin{figure}[!ht]
\centering
\subfigure[center][{ISD=500 meters.}]{
\includegraphics[width=0.45\columnwidth]{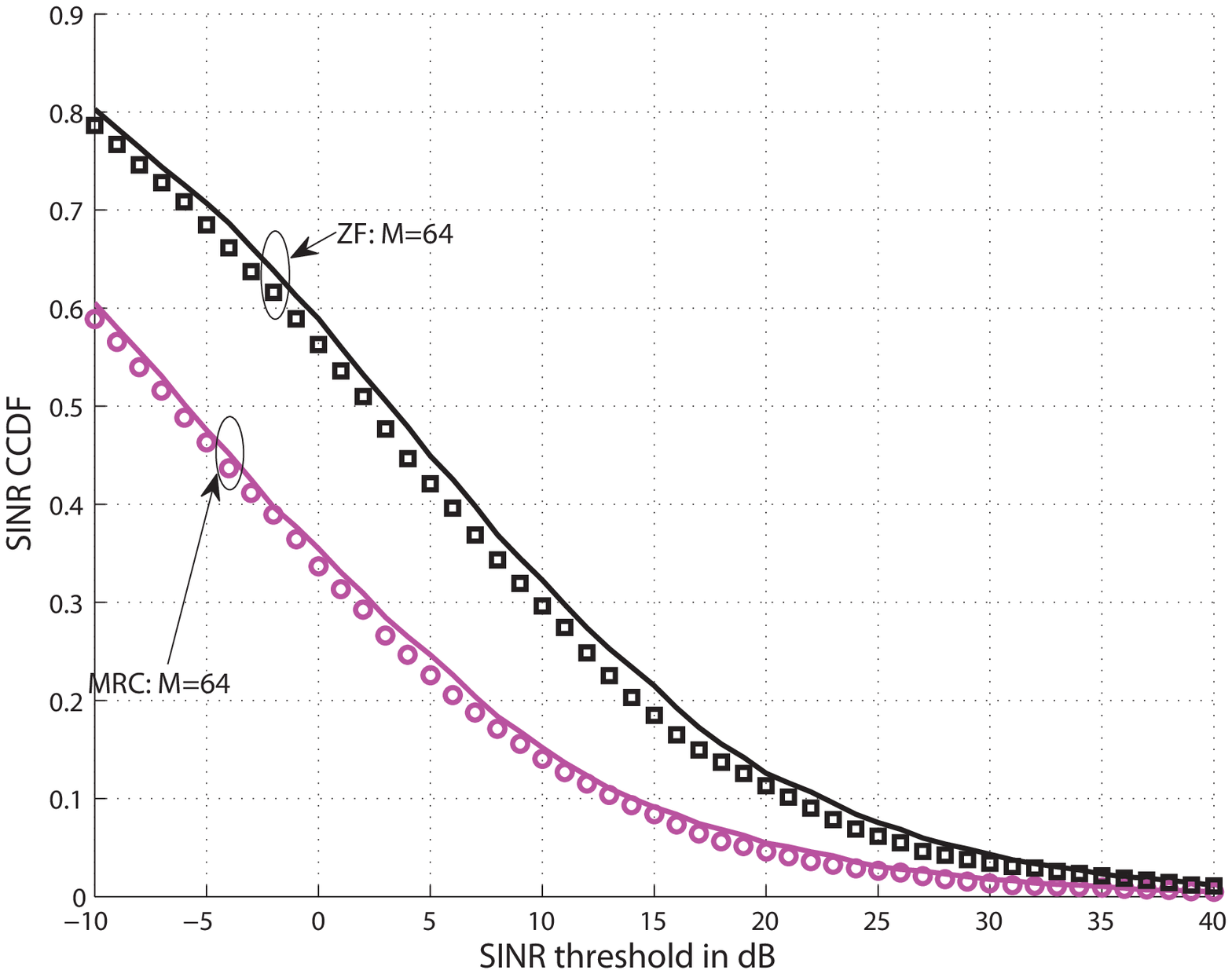}}
\subfigure[center][{ISD=1000 meters.}]{
\includegraphics[width=0.45\columnwidth]{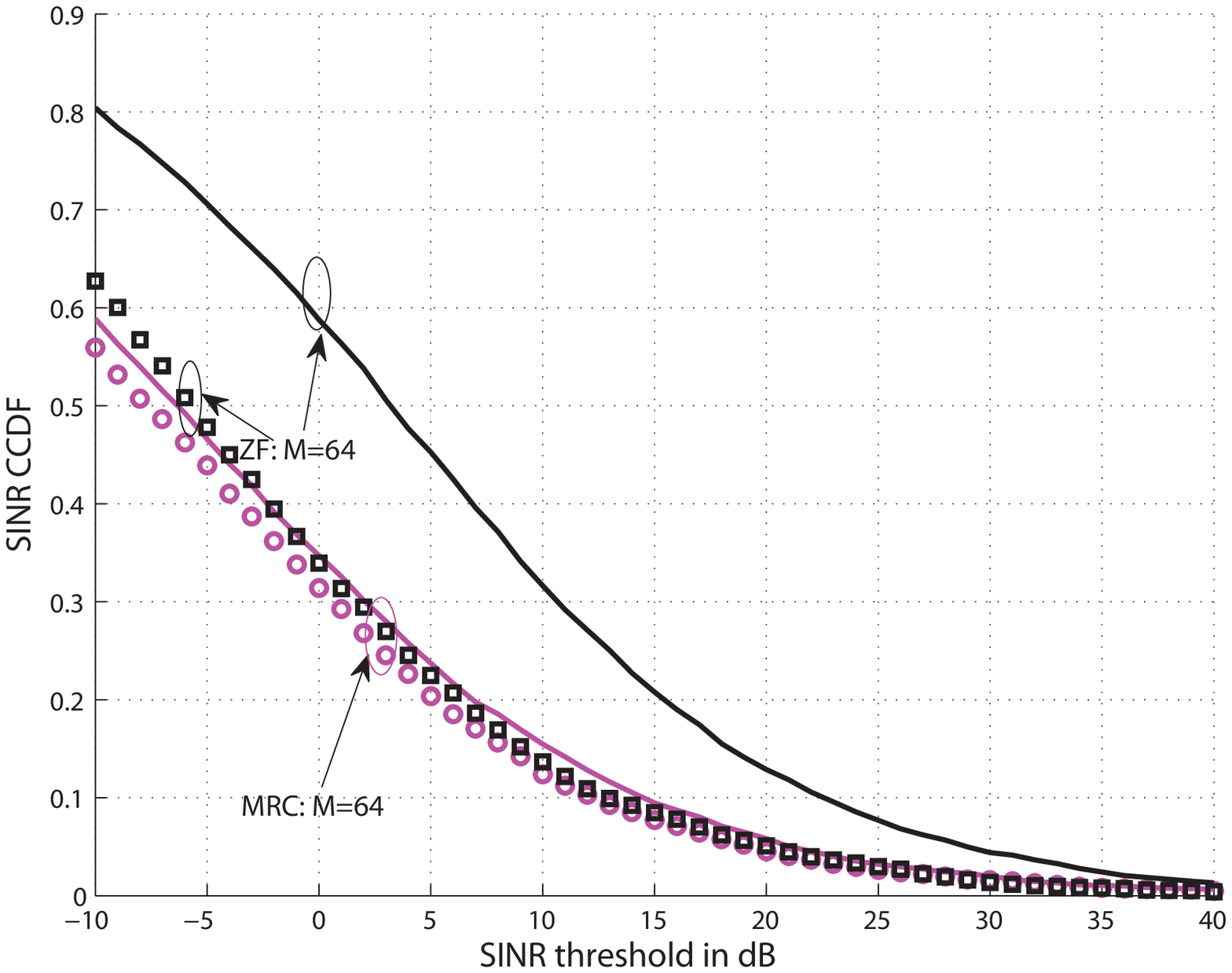}}
\caption{Comparison of SINR and SIR distributions. In the figures, we use markers to represent SINR curves, solid lines for SIR. We assume $K=10$ users per cell, $\epsilon=0$, and $\alpha=4$ in all cases. The gap between the SIR and SINR distributions becomes minor when ISD=500 meters, which is the typical size for the urban macro cells \cite{3GPPTR36.8142010}.}\label{fig:noise}
\end{figure}

\begin{figure} [!ht]
\centerline{
\includegraphics[width=0.45\columnwidth]{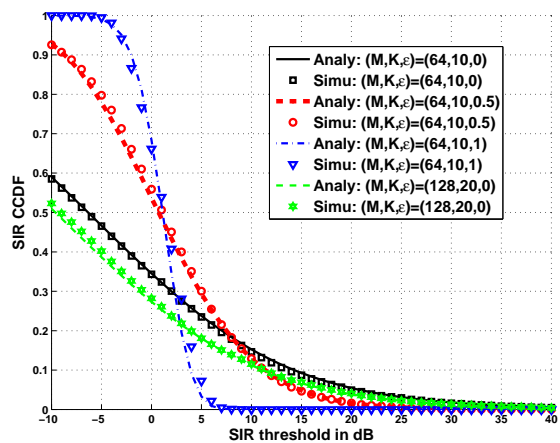}
   }
\caption{SIR coverage for \ac{MRC} receivers. In the simulations, we assume $\alpha=4$. The analytical curves are drawn based on Theorem \ref{thm:MRC_SIR}, which are shown a good fit with simulation. The difference in the curves for $(M,K,\epsilon)=(64,10,0)$ and $(M,K,\epsilon)=(128,20,0)$ indicates that linear scaling between $M$ and $K$ does not generally preserve SIR for \ac{MRC} receivers.}\label{fig:MRC_coverage}
\end{figure}


\begin{figure} [!ht]
\centerline{
\includegraphics[width=0.45\columnwidth]{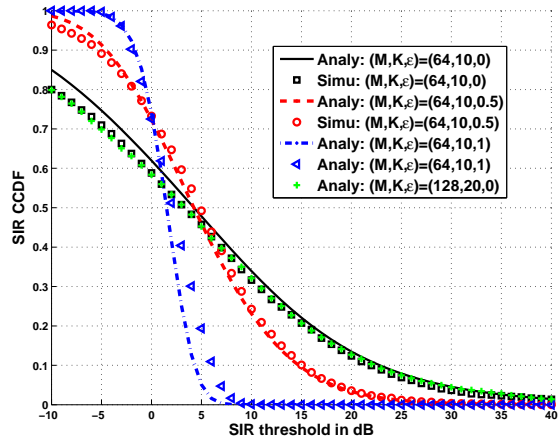}
   }
\caption{SIR distributions with \ac{ZF} receivers. We assume $\alpha=4$, and \ac{IID} fading channel. The analytical curves are plotted based on Theorem \ref{thm:ZF_SIR}. Simulations verify the analytical results, and show that when both $M$ and $K$ double, the SIR curves remain almost unchanged.}\label{fig:ZF_coverage}
\end{figure}

\begin{figure} [!ht]
\centerline{
\includegraphics[width=0.45\columnwidth]{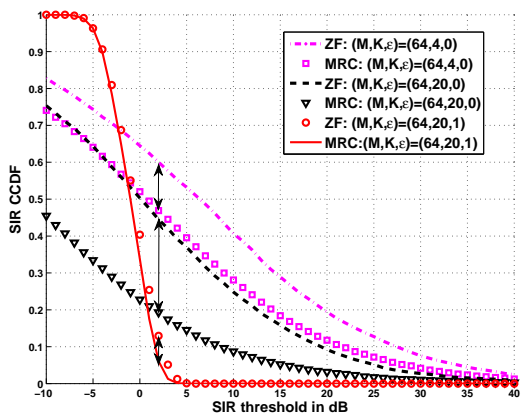}
   }
\caption{Comparison of SIR coverage with \ac{MRC} and \ac{ZF} receivers. We assume $\alpha=4$. As the double arrays display, when fixing $\epsilon=0$, the performance gap in SIR coverage is shown to increase with $K$; when fixing $K=20$, the gap diminishes when $\epsilon\to1$.}\label{fig:ZF-MRC_a}
\end{figure}

\begin{figure} [!ht]
\centerline{
\includegraphics[width=0.45\columnwidth]{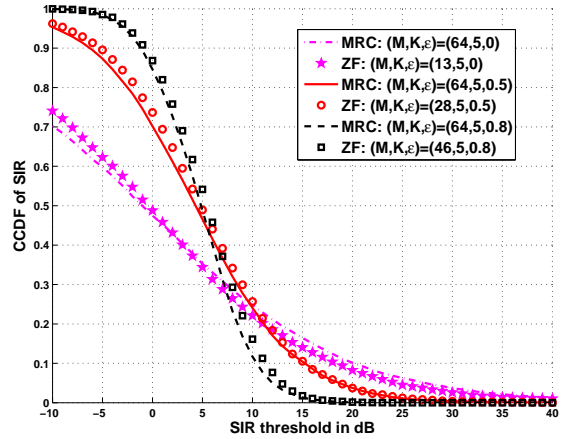}
   }
\caption{Verification of Proposition \ref{thm:comparison}. In the simulation, $\alpha=4$. In the simulation, we use the SIR curve of $\mathrm{MRC}(64,5)$ as a baseline for comparison. We use Proposition \ref{thm:comparison} to compute the required number of antennas for \ac{ZF} receivers, to have the SIR distribution of the baseline curve.}\label{fig:ZF_MRC_b}
\end{figure}

%

\begin{figure}[!ht]
\centering
\subfigure[center][{SIR for \ac{MRC} receivers in the hexagonal grid model.}]{
\includegraphics[width=0.45\columnwidth]{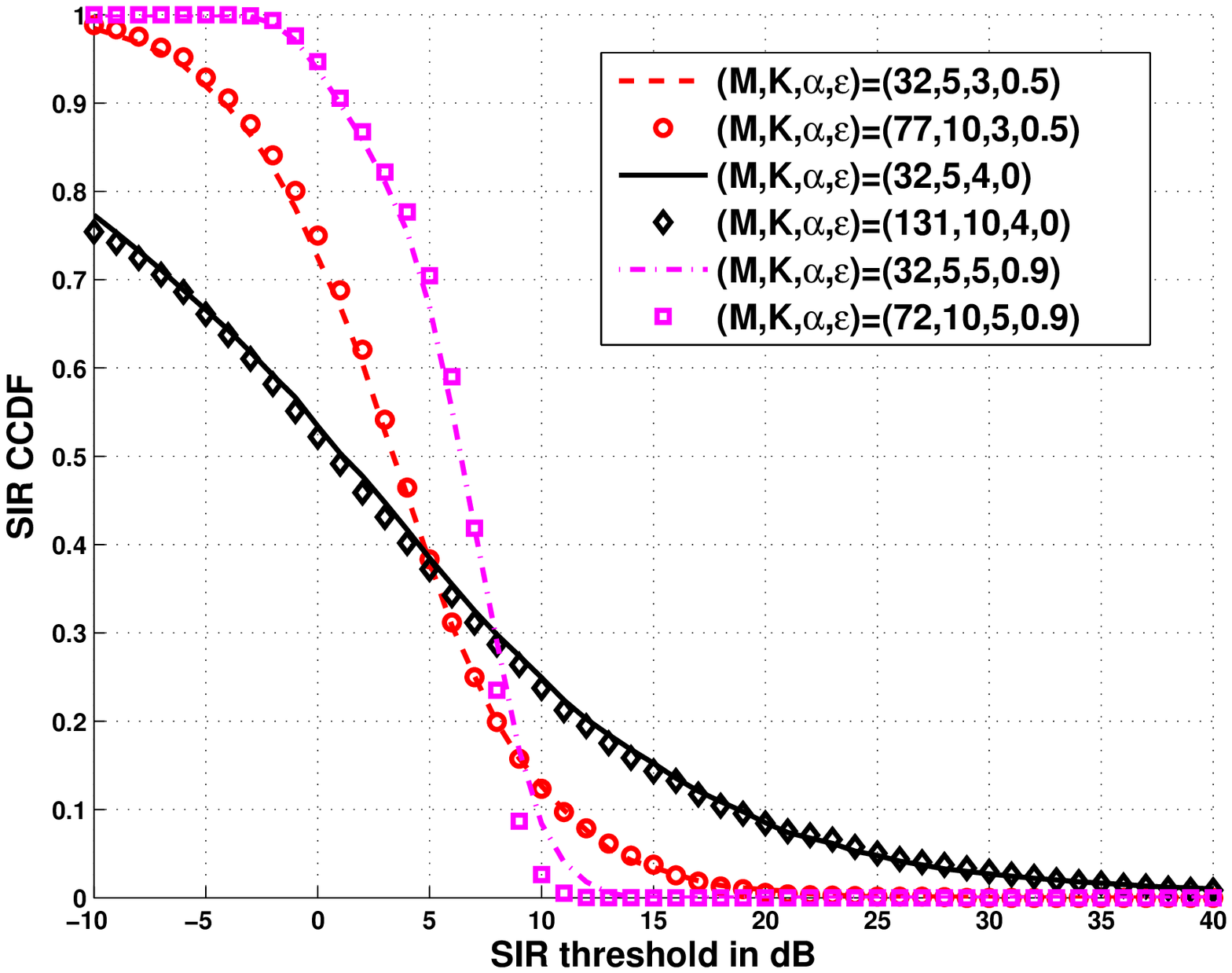}}
\subfigure[center][{SIR for \ac{ZF} receivers in the hexagonal grid model.}]{
\includegraphics[width=0.45\columnwidth]{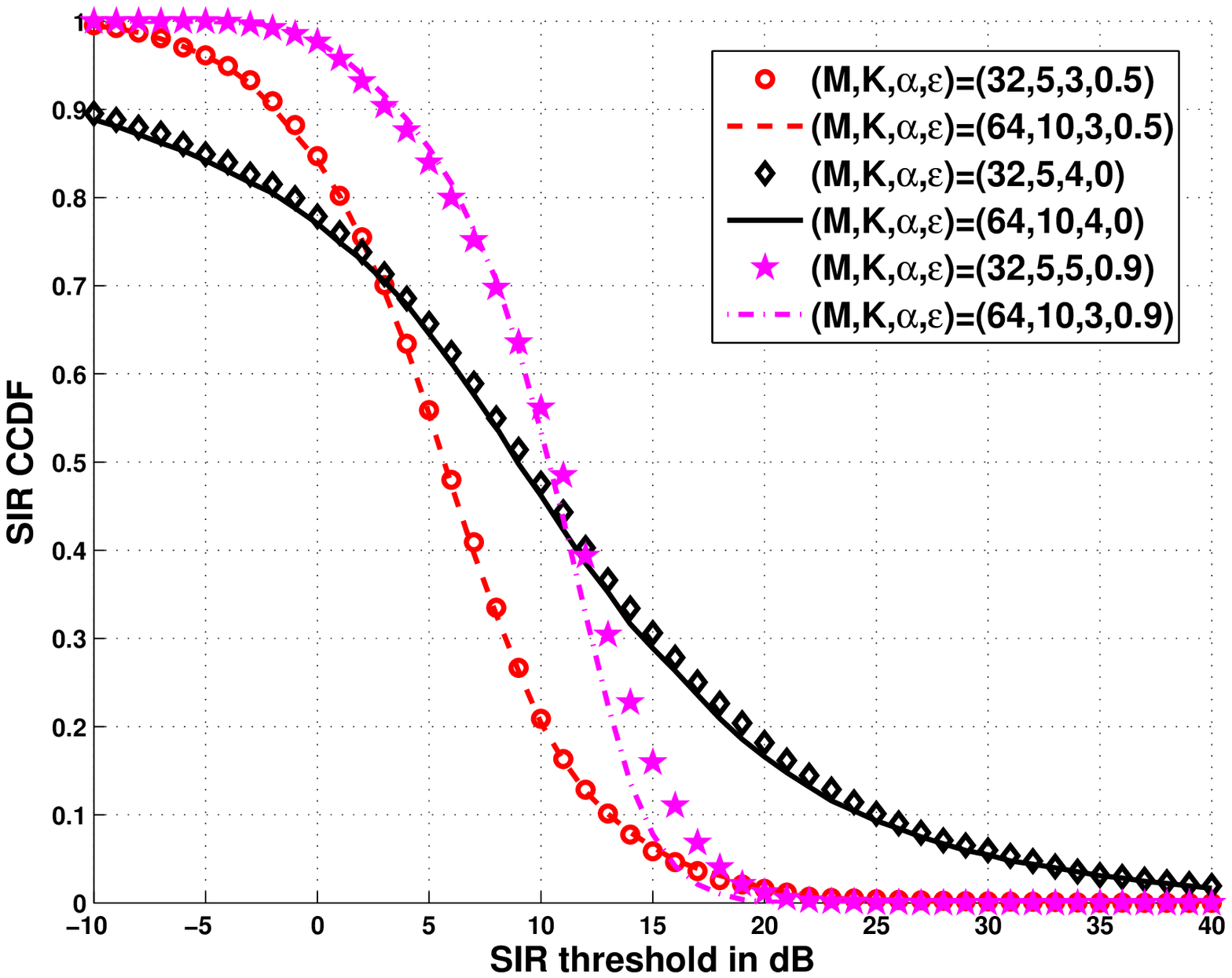}}
\caption{Verification of the scaling laws in the hexagonal model. We use $(M,K)=(32,5)$ as the baseline curves. When increasing the number of users to $K$=10, we compute the required $M$ to preserve the SIR distribution as baseline curves, according to Scaling law \ref{cor:conj} and Scaling law \ref{lem:scaling_ZF}. Simulations indicates that the scaling law results apply to the hexagonal model.}\label{fig:hex_simu_new}
\end{figure}

\begin{figure} [!ht]
\centerline{
\includegraphics[width=0.45\columnwidth]{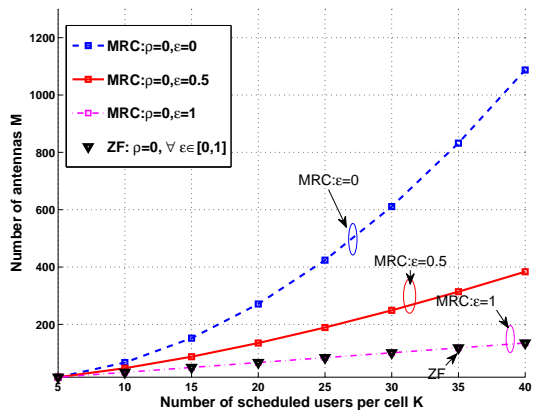}
   }
\caption{Comparison of different scaling laws. We plot the required number of antennas to provide the same SIR as that of the case $(M,K)=(16,5)$ as a function of $K$ with different system parameters.}\label{fig:scaling_law}
\end{figure}

\begin{figure} [!ht]
\centerline{
\includegraphics[width=0.45\columnwidth]{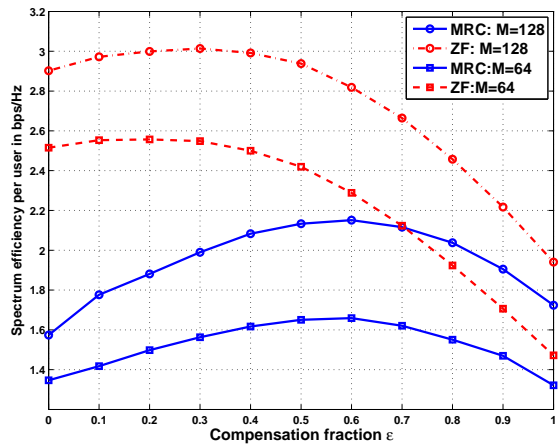}
   }
\caption{Average spectrum efficiency per user in an interference-limited network. In the simulation, we assume $T_{\max}=21$ dB, which sets the maximum spectrum efficiency per data stream as 7 bps/Hz. Training overhead is not taken account in this figure.}\label{fig:per_user_rate}
\end{figure}

\begin{figure}[!ht]
\centering
\subfigure[center][{Cell throughput when $T_c$=40 symbols.}]{
\includegraphics[width=0.45\columnwidth]{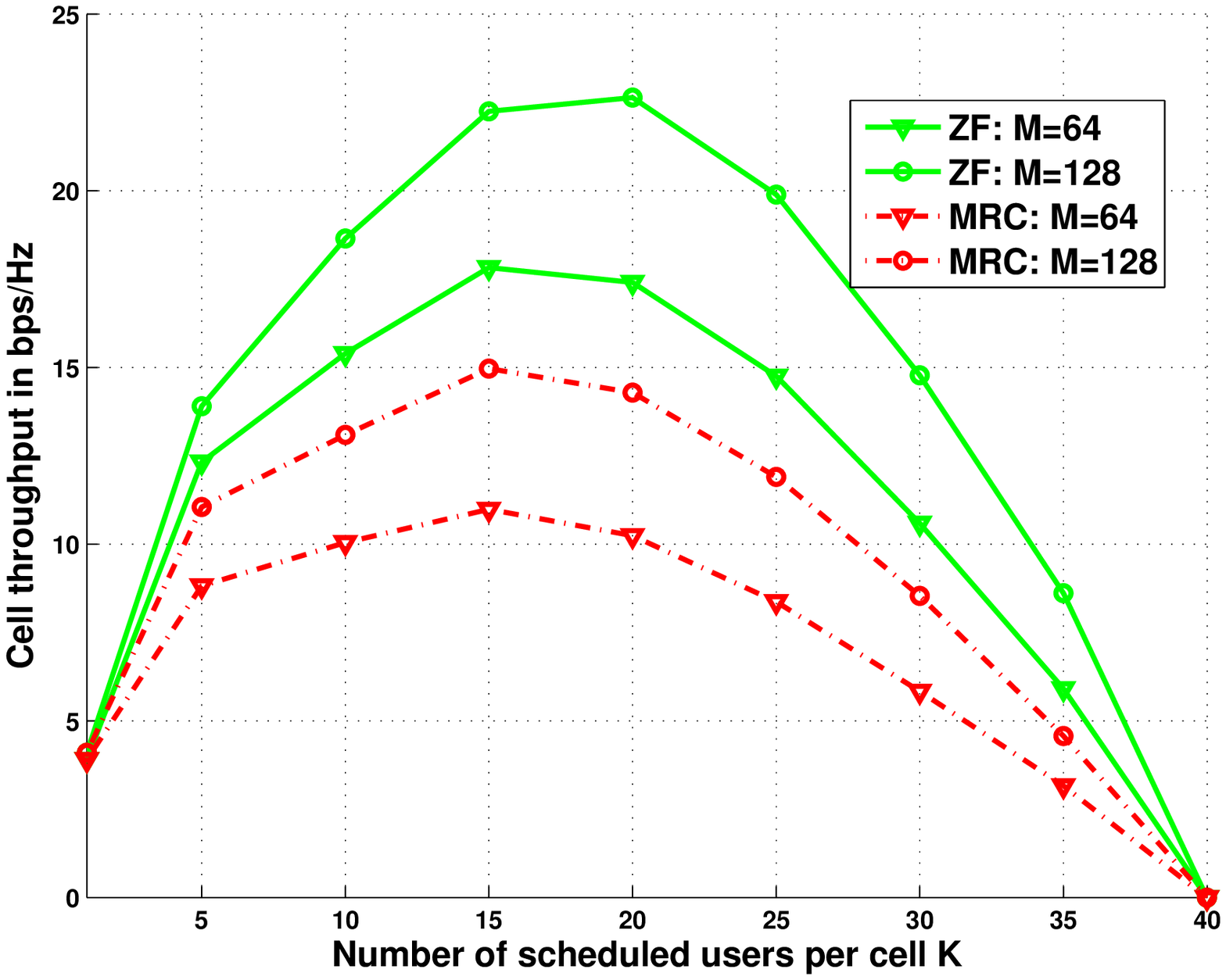}}
\subfigure[center][{Cell throughput when $T_c$=200 symbols.}]{
\includegraphics[width=0.45\columnwidth]{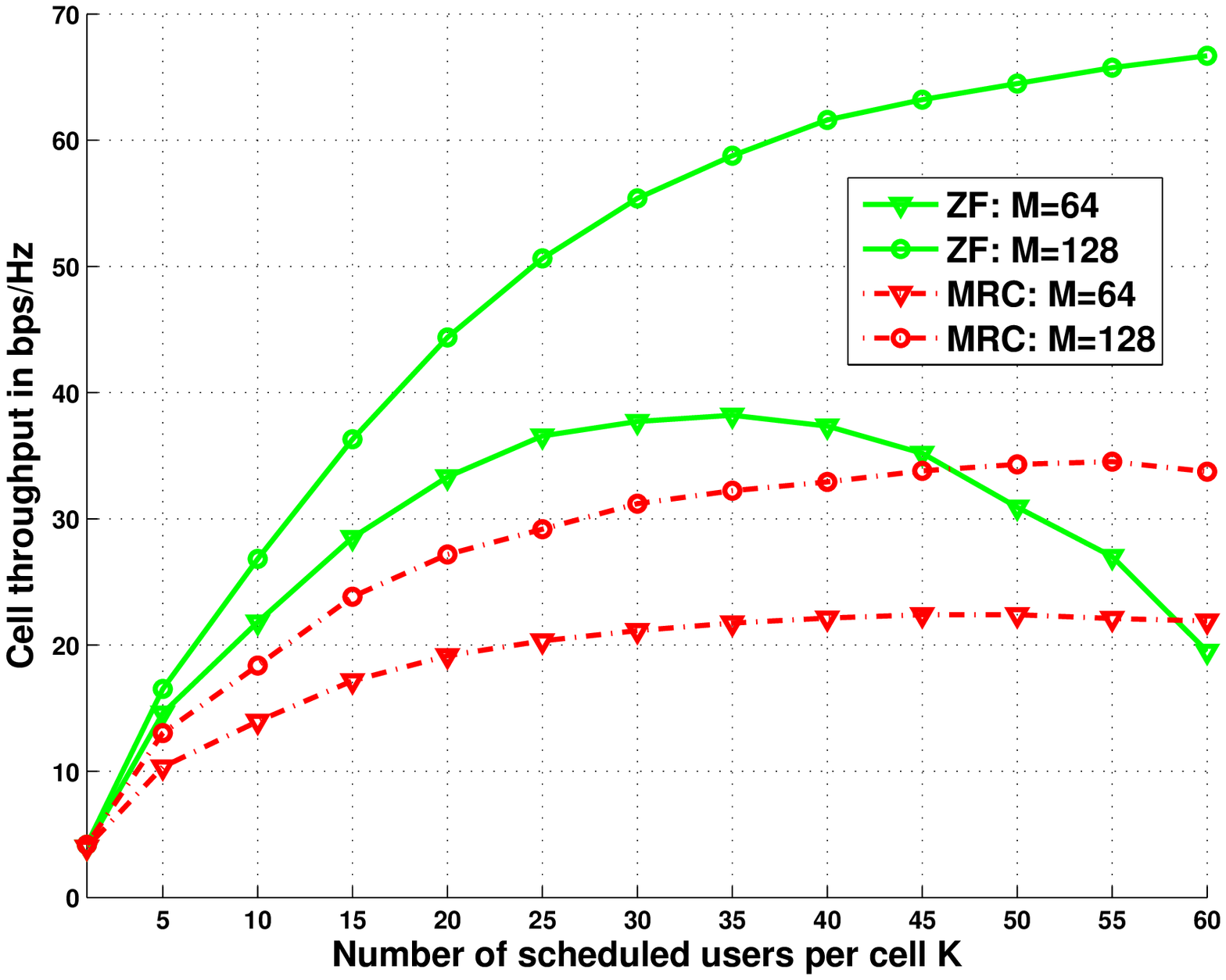}}
\caption{Uplink cell throughput as a function of $K$. The overhead due to channel training is taken account in the simulations. We simulate an interference-limited network with ISD=500 meters. We use $\epsilon=0.5$ for \ac{MRC} receivers, and $\epsilon=0.2$ for \ac{ZF} receivers, which are shown to optimize the per user rate. }\label{fig:sum_rate}
\end{figure}

{\bf Impact of the noise:} To begin with, we examine the impact of noise by comparing the SINR and SIR distributions for both \ac{MRC} and \ac{ZF} receivers in different scenarios in Fig. \ref{fig:noise}. In the simulations, we assume $P_\mathrm{t}$=23 dBm, and the bandwidth is 20 MHz as in the current LTE standards \cite{3GPPTR36.8142010}. We examine the case of $\epsilon=0$, which maximizes the impact of the noise. We simulate with two inter-site distances (ISDs): an average ISD of 500 meters in Fig. \ref{fig:noise}(a), and 1000 meters in Fig. \ref{fig:noise}(b). Note that a typical ISD of 500 meters is assumed for urban macro-cells in the 3GPP standards \cite{3GPPTR36.8142010}. In Fig. \ref{fig:noise}(a), the network with ISD=500 meters is shown to be interference-limited with $M=64$ antennas for both \ac{MRC} and \ac{ZF} receivers, as the SIR curves almost coincide with the SINR curves, which justifies the interference-limited assumption in urban marco cells. In the sparse network with ISD=1000 meters, however, simulations show that even with $M$=64 antennas, notable gaps exist between the SINR and SIR distributions, especially for \ac{ZF} receivers. In addition, the results in Fig. \ref{fig:noise}(b) shows that when the noise power is high, even with no power control, \ac{ZF} and \ac{MRC} receivers have a similar SINR coverage performance, which indicates that the SIR comparison results in Proposition \ref{thm:comparison} need not extends to general SINR comparisons.

 {\bf SIR coverage for \ac{MRC} receivers:} In Fig. \ref{fig:MRC_coverage}, we verify the analytical results for the \ac{SIR} distribution with \ac{MRC} receivers. Numerical results show that the SIR coverage is sensitive to the compensation fraction $\epsilon$ in the fractional power control: a large compensation fraction $\epsilon$ improves the SIR coverage in the low SIR regime at the expense of sacrificing the coverage in the high SIR regime. Besides, a comparison of the curves for $(M,K)=(64,10)$ and $(M,K)=(128,20)$ shows that the linear scaling law does not maintain the SIR distribution when $\epsilon=0$.


{\bf SIR coverage for \ac{ZF} receivers:} We verify the analysis for \ac{ZF} receivers in Fig. \ref{fig:ZF_coverage}. The analytical curves generally match well with numerical simulations. A comparison of the curves for $(M,K,\epsilon)=(64,10,0)$ and $(M,K,\epsilon)=(128,20,0)$ shows that unlike the case of \ac{MRC} receivers, a linear scaling law between $M$ and $K$ maintains the SIR distribution, even when there is no fractional power control implemented.

{\bf SIR comparison between \ac{MRC} and \ac{ZF} receivers:} We compare the uplink \ac{SIR} distributions for \ac{MRC} and \ac{ZF} receivers in Fig. \ref{fig:ZF-MRC_a}. Simulations show that \ac{ZF} receivers provide better \ac{SIR} coverage, due to the suppression of intra-cell interference. Moreover, for the same $\epsilon$, the performance gap between \ac{MRC} and \ac{ZF} receivers increases with the number of scheduled users $K$, as the strength of total intra-cell interference also increases with $K$. When fixing $M$ and $K$, the performance gap decreases with $\epsilon$; when $\epsilon=1$, the SIR coverage gap becomes minimal between \ac{MRC} and \ac{ZF} receivers. With full compensation of path loss in power control, linear scaling laws between $M$ and $K$ apply to both \ac{MRC} and \ac{ZF} receivers, as the near-far effect for users in a cell is mitigated. When $M\gg K$, the difference in the average (residue) intra-cell interference between \ac{MRC} and \ac{ZF} receivers becomes minor, as it decays with $\frac{1}{M}$.

In Fig. \ref{fig:ZF_MRC_b}, we verify our theoretical results in Proposition \ref{thm:comparison}. In the simulation, we fix the number of antennas for the \ac{MRC} receivers to be $M_\mathrm{MRC}=64$, and use Proposition \ref{thm:comparison} to calculate the required $M_\mathrm{ZF}$, to maintain the same SIR distribution. Numerical results show a good match with our analysis; the minor mismatch in the case $\epsilon=0$ is because Proposition \ref{thm:comparison} theoretically requires $\frac{M_\mathrm{ZF}}{K}\gg1$, while we use $\frac{M_\mathrm{ZF}}{K}=\frac{13}{5}$ in the simulation.

{\bf Verification with hexagonal grid model:} We verify the scaling laws derived from stochastic geometry with the hexagonal grid model in Fig. \ref{fig:hex_simu_new}. In the simulations, we use a layout of 19 hexagonal cells with inter-site distance of $300$ meters; only the scheduled users in the central cell are counted for the SIR statistics, to avoid edge effect. In Fig. \ref{fig:hex_simu_new}(a), for \ac{MRC} receivers, we use a $(M,K)=(32,5)$ as the baseline curve for comparison. When doubling the number of scheduled users to $K=10$, we use Scaling law \ref{cor:conj} to compute the required $M$ to maintain the same SIR distribution, which is shown to be almost accurate with extensive combinations of the system parameters in the hexagonal grid model. Similarly, results in Fig. \ref{fig:hex_simu_new}(b) verifies the linear scaling law for \ac{ZF} receivers in Scaling law \ref{lem:scaling_ZF}. This indicates that the stochastic geometry model provides reasonable predictions even for the hexagonal model.

{\bf Comparison of scaling laws:}  We compare scaling laws to maintain the uplink SIR distribution in different scenarios in Fig. \ref{fig:scaling_law}. We plot the required number of antennas to maintain the same SIR distribution as that in the case of $(M,K)=(16,5)$, as a function of $K$. As shown in the plot, for \ac{MRC} receivers, given the path loss exponent $\alpha$, the slope of the scaling law is determined by the fraction of path loss compensation $\epsilon$: the linear scaling law proposed in prior work \cite{Liang2015,Hoydis2013} is only achieved when $\epsilon=1$. Although the choice of $\epsilon=1$ makes the system with \ac{MRC} receivers linearly scalable, it need not maximize the per-user rate, as will be shown in Fig. \ref{fig:per_user_rate}. On the contrary, for \ac{ZF} receivers, the linear scaling applies for all $\epsilon\in[0,1]$.

\begin{table}
\begin{center}
\caption{Coherence time $T_c$ in the examples}\label{table:doppler}
\begin{tabular}{ c|c | c | c }
\hline
   \mbox{Mobility} &\mbox{Max. velocity} & \mbox{Max. Doppler} $f_\mathrm{D}$ &\mbox{Coherence time} $T_\mathrm{c}$\\ \hline
   \mbox{High}&\mbox{50 Km/h} &\mbox{92.6 Hz} & \mbox{40 Symbols}\\ \hline
    \mbox{Low}&\mbox{10 Km/h} &{18.5 Hz}& \mbox{200 Symbols} \\ \hline
 \end{tabular}
  \end{center}
\end{table}
{\bf Rate performance:} We illustrate the results on the average spectrum efficiency per user in Fig. \ref{fig:per_user_rate}. In the simulation, the average ISD is 500 meters, and $K=10$, which is shown to be interference-limited in Fig. \ref{fig:noise}. Consistent with the SIR results, in a interference-limited network, \ac{ZF} receivers provide a higher spectrum efficiency per user. Numerical results also show that the average spectrum efficiency is sensitive to the fraction of the path loss compensation $\epsilon$; the optimum $\epsilon$ for per user rate is generally around 0.5 for \ac{MRC} receiver, and 0.2 for \ac{ZF} receivers. In addition, we also observe that there is a minor performance gap in rate between \ac{ZF} and \ac{MRC} receivers under full channel compensation power control, as predicted by Proposition \ref{thm:comparison}.

Last, we examine the cell throughput in a system operated at 2 GHz in Fig. \ref{fig:sum_rate}. As an example, we consider an OFDM system, where the symbol time is 66.7 $\mu s$. We consider two cases with different mobilities as listed in Table \ref{table:doppler}; the coherence time $T_\mathrm{c}$ is computed as $T_\mathrm{c}=\frac{1}{4f_\mathrm{D}}$ \cite{Tse2005}, where $f_\mathrm{D}$ is the maximum doppler frequency. In this simulation, we assume the density of overall users to be 100 times the base station density, to simulate the case with large $K$. In Fig. \ref{fig:sum_rate}(a), in the high mobility case, when $T_\mathrm{c}<M$, the optimal $K^*$ for cell throughput is limited by the duration of $T_\mathrm{c}$, and the optimal value generally is $K^*\le \frac{T_\mathrm{c}}{2}$. In the example of low mobility case, when $T_\mathrm{c}>M$, the results in Fig. \ref{fig:sum_rate}(b) show that the optimal $K^*$ depends much on $M$: for \ac{ZF} receivers, the cell throughput drops fast when $\frac{K}{M}$ approaches to 1, and the optimal $K^*$ is around $\frac{M}{2}$ for maximum throughput; for \ac{MRC} receivers, the cell throughput becomes saturated approximately when $K>\frac{M}{3}$. In addition, \ac{ZF} receivers generally achieve better cell throughput than \ac{MRC} receivers; the only exception is the case of $\frac{M}{K}\approx1$, where the cell throughput of \ac{ZF} receivers drops below that of \ac{MRC}. In addition, compared with the single user per cell case ($K=1$), the results confirm that massive MIMO improves the cell throughput by serving multiple users simultaneously.

\section{Conclusions}\label{sec:conclusion}
In this paper, we proposed a stochastic geometry framework to analyze the spatial average SINR coverage and rate in massive \ac{MIMO} networks. We applied the analysis and numerical results to draw several important system design insights about the SINR coverage and rate in uplink massive \ac{MIMO} networks.

\begin{itemize}
\item The uplink massive MIMO networks can be interference-limited in urban marco cells (ISD=500 meters) with $M=64$ antennas at base stations.
\item With \ac{MRC} receivers, the number of antennas $M$ should scale super-linearly with the number of scheduled users per cell $K$ as $(M+1)\sim K^{\frac{\alpha}{2}(1-\epsilon)+\epsilon}$, to maintain the uplink SIR distribution; a linear scaling law only applies to the case of full path loss compensation in the power control, i.e., when $\epsilon=1$.
\item With \ac{ZF} receivers, a linear scaling between the number of antennas $M$ and users per cell $K$ maintains the uplink SIR distribution in massive \ac{MIMO}.
\item When noise is negligible, \ac{ZF} receivers provide better \ac{SIR} coverage rate than \ac{MRC} receivers. The performance gap increases with $K$, and decreases with path loss compensation faction $\epsilon$. The gap becomes minor when $\epsilon=1$.
\item The SIR coverage and rate are sensitive to the fraction $\epsilon$ of path loss compensation in power control. Larger $\epsilon$ improves coverage in the low SIR regime while reducing coverage probability at high SIR. Numerical results show that the optimal $\epsilon$ for rate is around 0.5 for \ac{MRC}, and 0.2 for \ac{ZF} receivers in certain cases.
\end{itemize}


\section*{Appendix}

\subsection{Derivation of (\ref{eqn:SIR_MRC_simplified}):} With the combining vector in (\ref{eqn:mrc_bf}), the \ac{SINR} expression equals (\ref{eqn:SIR_MRC_approx}) as
\begin{align}\label{eqn:SIR_MRC_approx}
\mathrm{SIR}=\frac{P^{(1)}_{0}\left|\mathbf{u}_{00}^{(1)*}\bar{\mathbf{h}}_{00}^{(1)} \right|^2}{P^{(1)}_{0}\mathbb{E}\left|\mathbf{u}_{00}^{(1)*}\hat{\mathbf{h}}_{00}^{(1)} \right|^2+\sum_{(\ell,k)\ne(0,1)}P^{(k)}_{\ell}\mathbb{E}\left|\mathbf{n}_{\mathrm{t}}^{*}\mathbf{h}_{0\ell}^{(k)}+\sum_{\ell'\ge0}\sqrt{P^{(1)}_{\ell'}}\mathbf{h}_{0\ell'}^{(1)*}\mathbf h_{0\ell}^{(k)}\right|^2+|\mathbf{u}_{00}^{(1)}|^2\sigma^2}.
\end{align}
In the numerator, the signal power can be computed as
\begin{align*}
P^{(1)}_{0}|\mathbf{u}_{00}^{(1)*}\bar{\mathbf{h}}_{00}^{(1)} |^2
&\stackrel{(a)}=\frac{\left(P_0^{(1)}\beta_{00}^{(1)}\right)^2}{\left(\sum_{\ell}P_\ell^{(1)}\beta_{0\ell}^{(1)}+\frac{\sigma^2}{K}\right)^2}|\mathbf{u}_{00}^{(1)}|^4\stackrel{(b)}\approx\frac{\left(P_0^{(1)}\beta_{00}^{(1)}\right)^2}{\left(\sum_{\ell}P_\ell^{(1)}\beta_{0\ell}^{(1)}+\frac{\sigma^2}{K}\right)^2}\mathbb{E}|\mathbf{u}_{00}^{(1)}|^4\\
&\stackrel{(c)}=\left(P_0^{(1)}\beta_{00}^{(1)}\right)^2(M^2+M)=P_\mathrm{t}^2\left(\beta_{00}^{(1)}\right)^{2(1-\epsilon)}(M^2+M),
\end{align*}
where (a) follows from the MMSE estimator in (\ref{eqn:MMSE1}), (b) follows from the fact that $|\mathbf{u}_{00}^{(1)}|^4\stackrel{M\to\infty}\to\mathbb{E}|\mathbf{u}_{00}^{(1)}|^4$, and the approximation error decays as $\frac{1}{M^2}$\cite{Ngo2014}, and (c) follows from the fact that $\mathbb{E}|\mathbf{u}_{00}^{(1)}|^4=(M^2+M)\left(\sum_{\ell}P_\ell^{(1)}\beta_{0\ell}^{(1)}+\frac{\sigma^2}{K}\right)^2$.
Next, we compute the first term in the denominator of (\ref{eqn:SIR_MRC_approx}) as
\begin{align*}
P^{(1)}_{0}\mathbb{E}|\mathbf{u}_{00}^{(1)*}\hat{\mathbf{h}}_{00}^{(1)} |^2&=\frac{1}{M}P^{(1)}_{0}\mathbb{E}|\mathbf{u}_{00}^{(1)}|^2\mathbb{E}|\hat{\mathbf{h}}_{00}^{(1)}|^2\\
&\stackrel{(a)}=\frac{1}{M}P^{(1)}_{0}M\left(\sum_{\ell}P^{(1)}_{\ell}\beta^{(1)}_{0\ell}+\frac{\sigma^2}{K}\right)M\beta_{00}^{(1)}\left(1-\frac{P^{(1)}_{0}\beta^{(1)}_{00}}{\sum_{\ell}P^{(1)}_{\ell}\beta^{(1)}_{0\ell}+\frac{\sigma^2}{K}}\right)\\
&=MP_\mathrm{t}^2\left(\beta_{00}^{(1)}\right)^{1-\epsilon}\left(\sum_{\ell>0}\left(\beta_{\ell\ell}^{(1)}\right)^{-\epsilon}\beta_{0\ell}^{(1)}+\frac{\sigma^2}{P_\mathrm{t}K}\right),
\end{align*}
where (a) follows from the fact that the channel estimation error $\hat{\mathbf{h}}_{00}^{(1)}$ follows the distribution $\mathcal{CN}\left(\mathbf{0},\beta_{00}^{(1)}\left(1-\frac{P^{(1)}_{0}\beta^{(1)}_{00}}{\sum_{\ell}P^{(1)}_{\ell}\beta^{(1)}_{0\ell}+\frac{\sigma^2}{K}}\right)\mathbf{I}_M\right)$.

Next, we simplify the second term in the denominator. Note that unless $(\ell,\ell',k)=(n,n',m)$, $\mathbf{h}_{\ell'0}^{(1)*}\mathbf h_{\ell0}^{(k)}$ and $\mathbf{h}_{n'0}^{(1)*}\mathbf h_{n0}^{(m)}$ are uncorrelated zero-mean random variables. Therefore, we can simplify the second term in the denominator of (\ref{eqn:SIR_MRC_approx}) as
\begin{align}\label{eqn:SIR_approx55}
&\sum_{(\ell,k)\ne(0,1)}\mathbb{E}|\mathbf{n}_\mathrm{t}^{*}\mathbf h_{\ell0}^{(k)}+\sum_{\ell'\ge0}\sqrt{P^{(1)}_{\ell'}}\mathbf{h}_{\ell'0}^{(1)*}\mathbf h_{\ell0}^{(k)}|^2\\
&=\sum_{(\ell,k)\ne(0,1)}\left(\frac{\sigma^2}{K}MP_\mathrm{t}\left(\beta_{\ell\ell}^{(k)}\right)^{-\epsilon}\beta_{\ell0}^{(k)}+\sum_{\ell'>0}P^{(k)}_{\ell}P^{(1)}_{\ell'}\mathbb{E}\left[|\mathbf{h}_{\ell'0}^{(1)*}\mathbf h_{\ell0}^{(k)}|^2\right]\right).\label{eqn:SIR_approx55}
\end{align}
Note that for $k=1$,$\ell'=\ell\ne0$, the expression is simplified as
\begin{align*}
P^{(k)}_{\ell}P^{(1)}_{\ell'}\mathbb{E}\left[|\mathbf{h}_{\ell'0}^{(1)*}\mathbf h_{\ell0}^{(k)}|^2\right]
&=P^{(1)2}_{\ell}\mathbb{E}\left[|\mathbf{h}_{\ell0}^{(1)}|^4\right]\\
&=(M^2+M)P^{(1)2}_{\ell}\left(\beta^{(1)}_{0\ell}\right)^2=P_\mathrm{t}^2(M^2+M)\left(\beta^{(1)}_{\ell\ell}\right)^{-2\epsilon}\left(\beta^{(1)}_{0\ell}\right)^2;
\end{align*}
for $k>1$ or $k=1$, $\ell'\ne\ell>0$, it follows that
\begin{align*}
P^{(k)}_{\ell}P^{(1)}_{\ell'}\mathbb{E}\left[|\mathbf{h}_{\ell'0}^{(1)*}\mathbf h_{\ell0}^{(k)}|^2\right]&=M P^{(1)}_{\ell'}P^{(k)}_{\ell}\beta^{(1)}_{0\ell'}\beta^{(k)}_{0\ell}\\&=M P_{\mathrm{t}}^2 \left(\beta^{(1)}_{\ell'\ell'}\beta^{(k)}_{\ell\ell}\right)^{-\epsilon}\beta^{(1)}_{0\ell'}\beta^{(k)}_{0\ell}.
\end{align*}
Therefore, we can express (\ref{eqn:SIR_approx55}) as
\begin{align*}
&\sum_{(\ell,k)\ne(0,1)}\left(\frac{\sigma^2}{K}MP_\mathrm{t}\left(\beta_{\ell\ell}^{(k)}\right)^{-\epsilon}\beta_{\ell0}^{(k)}+\sum_{\ell'>0}P^{(k)}_{\ell}P^{(1)}_{\ell'}\mathbb{E}\left[|\mathbf{h}_{\ell'0}^{(1)*}\mathbf h_{\ell0}^{(k)}|^2\right]\right)\\&=MP_\mathrm{t}^2\left(\sum_{(k,\ell)\ne(1,0)}\left(\beta_{\ell\ell}^{(k)}\right)^{-\epsilon}\beta_{\ell0}^{(k)}\right)\left(\frac{\sigma^2}{KP_\mathrm{t}}+\sum_{\ell'\ge0}\left(\beta^{(1)}_{\ell'\ell'}\right)^{-\epsilon}\beta^{(1)}_{0\ell'}\right)+M^2P^{2}_\mathrm{t}\sum_{\ell>0}\left(\beta^{(1)}_{\ell\ell}\right)^{-2\epsilon}\left(\beta^{(1)}_{0\ell}\right)^2 .
\end{align*}

Last, the thermal noise term in the denominator can be simplified as $$|\mathbf{u}_{00}^{(1)}|^2\sigma^2=P_\mathrm{t}M\sigma^2\left(\sum_{\ell}\left(\beta^{(1)}_{\ell\ell}\right)^{-\epsilon}\beta^{(1)}_{0\ell}+\frac{\sigma^2}{P_\mathrm{t}K}\right).$$

Then the expression in (\ref{eqn:SIR_MRC_simplified}) is obtained through algebraic manipulation in the denominator.

\subsection{Proof of Theorem \ref{thm:MRC_SIR}:}
To allow for tractable computation and decouple the correlated terms in the denominator of (\ref{eqn:SIR_MRC_simplified}), we propose the following approximations on the out-of-cell interference terms: for $k\in[1,K]$,
\begin{align}\label{eqn:approx_new1}
\Delta_1^{(k)}&\stackrel{(a)}\approx\mathbb{E}\left(\sum_{\ell>0}\left(\beta^{(k)}_{\ell\ell}\right)^{-\epsilon}\beta^{(k)}_{0\ell}\right)+\frac{\sigma^2}{KP_\mathrm{t}}=\mathbb{E}\left(\sum_{\ell>0}\mathbb{E}\left(\beta^{(k)}_{\ell\ell}\right)^{-\epsilon}\beta^{(k)}_{0\ell}\right)+\frac{\sigma^2}{KP_\mathrm{t}}\nonumber\\
&\stackrel{(b)}\approx C^{-\epsilon}(\lambda_\mathrm{b}\pi)^{-\frac{\alpha\epsilon}{2}}\Gamma^{\alpha}(\frac{\epsilon}{2}+1)\mathbb{E}\left(\sum_{\ell>0}\beta^{(k)}_{0\ell}\right)+\frac{\sigma^2}{KP_\mathrm{t}}\nonumber\\
&\stackrel{(c)}=\frac{2C^{1-\epsilon}(\lambda_\mathrm{b}\pi)^{\frac{\alpha(1-\epsilon)}{2}}}{\alpha-2}\Gamma^{\alpha}\left(\frac{\epsilon}{2}+1\right)+\frac{\sigma^2}{KP_\mathrm{t}},
\end{align}
where in (a), we approximate $\Delta_1^{(k)}$ by its mean; step (b) follows from
\begin{align}\label{eqn:SIR_approx1}
\mathbb{E}\left(\beta^{(k)}_{\ell\ell}\right)^{-\epsilon}=C^{-\epsilon}\left(\mathbb{E}\left[R^{(k)\epsilon}_{\ell\ell}\right]\right)^\alpha=(\lambda_\mathrm{b}\pi)^{-\frac{\alpha\epsilon}{2}}\Gamma^{\alpha}\left(\frac{\epsilon}{2}+1\right);
\end{align}
and step (c) follows from the exclusion ball model in Approximation \ref{approx:1} and the Campbell's theorem \cite{Baccelli2009a} as
\begin{align}\label{eqn:SIR_approx3}
\mathbb{E}\left[\sum_{\ell>0}\beta^{(k)}_{0\ell}\right]=2\pi\lambda_\mathrm{b}C\int_{R_\mathrm{e}}^{\infty}x^{-\alpha}x\mathrm{d}x=\frac{2C(\lambda_\mathrm{b}\pi)^{\alpha/2}}{\alpha-2}.
\end{align}
Similarly, we can approximate $\Delta_2^{(1)}$ by its mean as:
\begin{align}\label{eqn:approx_new2}
\Delta_2^{(k)}&\approx\frac{C^{2(1-\epsilon)}\left(\lambda_\mathrm{b}\pi\right)^{\alpha(1-\epsilon)}}{\alpha-1}\Gamma^{\alpha}(\epsilon+1),
\end{align}
where follows from
$
\mathbb{E}\left(\beta^{(k)}_{\ell\ell}\right)^{-2\epsilon}=C^{-2\epsilon}\left(\mathbb{E}\left[R^{(k)2\epsilon}_{\ell\ell}\right]\right)^\alpha=(\lambda_\mathrm{b}\pi)^{-\alpha\epsilon}\Gamma^{\alpha}(\epsilon+1).
$

Next, applying the approximation in (\ref{eqn:approx_new1}) and (\ref{eqn:approx_new2}) and conditioning on $R_{00}^{(1)}=x$, we simplify the approximate conditional uplink SINR as
{\small\begin{align}
\mathrm{SINR}\approx T\left(C_1(\lambda_\mathrm{b}\pi x^2)^{\alpha(1-\epsilon)}+C_2(\lambda_\mathrm{b}\pi x^2)^{\frac{\alpha}{2}(1-\epsilon)}+{C_4(\lambda_\mathrm{b}\pi x^2)}\sum_{k>0}\left(\lambda_\mathrm{b}\pi R^{(k)2}_{00}\right)^{-\frac{\alpha}{2}(1-\epsilon)}\right)^{-1},
\end{align}}
where in (a) $C_1$, $C_2$ and $C_4(x)$ are defined in Theorem \ref{thm:MRC_SIR}.
Next, conditioning on $R_{00}^{(1)}=x$, the approximate SINR distribution can be computed as
{\small\begin{align*}
&\mathbb{P}\left(\mathrm{SINR}>T|R_{00}^{(1)}=x\right)\\
&\stackrel{(a)}\approx\mathbb{P}\left(1>T\left(C_1(\lambda_\mathrm{b}\pi x^2)^{\alpha(1-\epsilon)}+C_2(\lambda_\mathrm{b}\pi x^2)^{\frac{\alpha}{2}(1-\epsilon)}+{C_4(\lambda_\mathrm{b}\pi x^2)}\sum_{k>0}\left(\lambda_\mathrm{b}\pi R^{(k)2}_{00}\right)^{-\frac{\alpha}{2}(1-\epsilon)}\right)\right)\\
&\stackrel{(b)}\approx\mathbb{P}\left(g>T\left(C_1(\lambda_\mathrm{b}\pi x^2)^{\alpha(1-\epsilon)}+C_2(\lambda_\mathrm{b}\pi x^2)^{\frac{\alpha}{2}(1-\epsilon)}+{C_4(\lambda_\mathrm{b}\pi x^2)}\sum_{k>0}\left(\lambda_\mathrm{b}\pi R^{(k)2}_{00}\right)^{-\frac{\alpha}{2}(1-\epsilon)}\right)\right)\\
&\stackrel{(c)}\approx 1-\mathbb{E}\left[\left(1-\mathrm{e}^{-\frac{\eta T}{M+1}\left(C_1(\lambda_\mathrm{b}\pi x^2)^{\alpha(1-\epsilon)}+C_2(\lambda_\mathrm{b}\pi x^2)^{\frac{\alpha}{2}(1-\epsilon)}+C_4(\lambda_\mathrm{b}\pi x^2)\sum_{k>0}\left(\lambda_\mathrm{b}\pi R^{(k)2}_{00}\right)^{-\frac{\alpha}{2}(1-\epsilon)}\right) }\right)^N\right]\\
&=\sum_{n=1}^{N}{{N}\choose{n}}(-1)^{n+1}\mathrm{e}^{-n \eta T\left(C_1\left(\lambda_\mathrm{b}\pi x^2\right)^{\alpha(1-\epsilon)}+C_2\left(\lambda_\mathrm{b}\pi x^2\right)^{\frac{\alpha}{2}(1-\epsilon)}\right)}\mathbb{E}\left[\mathrm{e}^{-n\eta T C_4\left(\lambda_\mathrm{b}\pi x^2\right)\left(\lambda_\mathrm{b}\pi R_{00}^{(k)2}\right)^{-\frac{\alpha}{2}(1-\epsilon)}}\right]^{K-1}\\
&\stackrel{(d)}=\sum_{n=1}^{N}{{N}\choose{n}}(-1)^{n+1}\mathrm{e}^{-n \eta T\left(C_1\left(\lambda_\mathrm{b}\pi x^2\right)^{\alpha(1-\epsilon)}+C_2\left(\lambda_\mathrm{b}\pi x^2\right)^{\frac{\alpha}{2}(1-\epsilon)}\right)}\left(\int_{0}^{\infty}\mathrm{e}^{-n\eta T C_4\left(\lambda_\mathrm{b}\pi x^2\right)s^{-{\alpha}(1-\epsilon)}-\lambda_\mathrm{b}\pi s^2}\lambda_\mathrm{b}\pi s\mathrm{d}s\right)^{K-1}\\
&\stackrel{(e)}=\sum_{n=1}^{N}{{N}\choose{n}}(-1)^{n+1}\mathrm{e}^{-n \eta T\left(C_1\left(\lambda_\mathrm{b}\pi x^2\right)^{\alpha(1-\epsilon)}+C_2\left(\lambda_\mathrm{b}\pi x^2\right)^{\frac{\alpha}{2}(1-\epsilon)}\right)}\left(\int_{0}^{\infty}\mathrm{e}^{-n\eta T C_4\left(\lambda_\mathrm{b}\pi x^2\right)u^{-\frac{\alpha}{2}(1-\epsilon)}-u}\mathrm{d}u\right)^{K-1}\\
&\stackrel{(f)}\approx\sum_{n=1}^{N}{{N}\choose{n}}(-1)^{n+1}\mathrm{e}^{-n \eta T\left(C_1\left(\lambda_\mathrm{b}\pi x^2\right)^{\alpha(1-\epsilon)}+C_2\left(\lambda_\mathrm{b}\pi x^2\right)^{\frac{\alpha}{2}(1-\epsilon)}\right)}\left(1-n \eta TC_4\left(\lambda_\mathrm{b}\pi x^2\right)\int_{0}^{\infty}\frac{\mathrm{e}^{-u}\mathrm{d}u}{1+n\eta T C_4\left(\lambda_\mathrm{b}\pi x^2\right)u^{-\frac{\alpha}{2}(1-\epsilon)}}\right)^{K-1},
\end{align*}
}where in (a) $C_1$, $C_2$ and $C_4(x)$ are defined in Theorem \ref{thm:MRC_SIR}; in (b) we use a ``dummy" gamma variable $g$ with unit mean and shape parameter $N$ to approximate the constant number one, and the approximation follows from the fact that $g$ converges to one when $N$ goes to infinity, i.e., $\lim_{n\to\infty}\frac{n^nx^{n-1}\mathrm{e}^{-nx}}{\Gamma(n)}=\delta(x-1)$ \cite{Aris1999}, where $\delta(x)$ is the Dirac delta function; in (c), the approximation follows from Alzer's inequality \cite[Appendix A]{Alzer1997,Bai2014}, where $\eta=N(N!)^{-\frac{1}{N}}$; (d) follows from the fact that $R_{00}^{(k)}$ is assumed to be \ac{IID} Rayleigh random variable; in (e) we change variable as $u=\lambda_\mathrm{b}\pi s^2$; in (f), we apply the approximation $\exp(-x)\approx \frac{1}{1+x}$ inside the integral, to allow for faster numerical evaluations. Last, we obtain the uplink SIR distribution by de-conditioning on $R_{00}^{(1)}=x$, which is assumed to be a Rayleigh random variable with mean $0.5\sqrt{1/\lambda_\mathrm{b}}$, and changing the variable as $t=\pi\lambda_\mathrm{b}x^2$.
\endproof

\subsection{Proof of Theorem \ref{thm:ZF_SIR}}

Before proving the theorem, we present a useful lemma on the distribution of the combining vector $\mathbf{g}_{00}^{(1)}$ as follows.

\begin{lemma}[From \cite{Kiessling2003}]
The square norm of the combining vector $|\mathbf{g}_{00}^{(1)}|^2$ follows the distribution of $\left(\chi^2_{2(K-M+1)}\left(\sum_{\ell'}P^{(1)}_{\ell'}\beta^{(1)}_{\ell\ell'}+\frac{\sigma^2}{K}\right)\right)^{-1}$, where $\chi_{2(K-M+1)}^2$ represents a Chi-square random variables with $2(K-M+1)$ degrees of freedom. \label{lem:com_vec_ZF1}
\end{lemma}
Note that when $(M_K+1)\to\infty$, $\frac{\chi_{2(K-M+1)}^2}{M-K+1}\to1$. Therefore, by Lemma \ref{lem:com_vec_ZF1}, when $M\gg K$, the following approximation holds as
 \begin{align}\label{eqn:ZF_approx}
|\mathbf{g}_{00}^{(1)}|^2\approx\frac{1}{(M-K+1)S^{(1)}},
\end{align}
where for ease of notation, we define $S^{(k)}=\sum_{\ell'}P^{(1)}_{\ell'}\beta^{(1)}_{\ell\ell'}+\frac{\sigma^2}{K}=\left(\beta_{00}^{(k)}\right)^{1-\epsilon}+\Delta_1^{(k)}$, and the approximation error decays as $\frac{1}{M-K+1}$.
Noting that $|\mathbf{g}_{00}^{(1)}\mathbf{u}_{00}^{(1)}|^2=1$, the nominator of the SINR expression in (\ref{eqn:UHF_UL_SIR}) can be computed as
\begin{align}
P^{(1)}_{0}|\mathbf{g}_{00}^{(1)*}\bar{\mathbf{h}}_{00}^{(1)} |^2&=\frac{P^{(1)2}_{0}\beta^{(1)2}_{00}}{\left(S^{(1)}\right)^2}|\mathbf{g}_{00}^{(1)}\mathbf{u}_{00}^{(1)}|^2=
\left(\beta_{00}^{(1)}\right)^{2(1-\epsilon)}/\left(S^{(1)}\right)^2.
\end{align}

Applying the results in (\ref{eqn:ZF_approx}), the first term in the denominator of (\ref{eqn:UHF_UL_SIR}) is computed as
\begin{align}\label{eqn:ZF_approx_1}
P^{(1)}_{0}\mathbb{E}|\mathbf{g}_{00}^{(1)*}\hat{\mathbf{h}}_{00}^{(1)} |^2&\approx P^{(1)}_{0}\beta_{00}^{(1)}\left(1-\frac{P^{(1)}_{0}\beta^{(1)}_{00}}{S^{(1)}}\right)\frac{1}{(M-K+1)S^{(1)}}\nonumber\\
&=\frac{\Delta_1^{(1)}\left(\beta_{00}^{(1)}\right)^{1-\epsilon}}{(M-K+1)\left(S^{(1)}\right)^2}.
\end{align}

Now we simplify the second sum in the denominator of (\ref{eqn:UHF_UL_SIR}): for $k=1$, and $\ell>0$, it follows
\begin{align}\label{eqn:ZF_approx_2}
\mathbb{E}|\mathbf{g}_{00}^{(1)*}\mathbf h_{0\ell}^{(1)}|^2&=\mathbb{E}|\mathbf{u}_{00}^{(1)}|^{-4}|\mathbf{g}_{00}^{(1)*}\mathbf u_{00}^{(1)}\mathbf u_{00}^{(1)*}\mathbf h_{\ell0}^{(1)}|^2=\mathbb{E}|\mathbf{u}_{00}^{(1)}|^{-4}|\mathbf u_{00}^{(1)*}\mathbf h_{\ell0}^{(1)}|^2\nonumber\\
&=\frac{(M^2+M)P_{\ell}^{(1)2}\beta_{0\ell}^{(1)2}+M\sum_{\ell'\ne\ell}P_{\ell}^{(1)}P_{\ell'}^{(1)}\beta_{0\ell}^{(1)}\beta_{0\ell'}^{(1)}+\frac{M\sigma^2}{K}P^{(1)}_{\ell}\beta^{(1)}_{0\ell}}{(M^2+M)\left(S^{(1)}\right)^2};
\end{align}
For $k>1$, it follows that
\begin{align}
&\mathbb{E}|\mathbf{g}_{00}^{(1)*}\mathbf h_{0\ell}^{(k)}|^2=\frac{\mathbb{E}|\mathbf{g}_{00}^{(1)}|^{2}\mathbb{E}\left(|\mathbf h_{0\ell}^{(k)}|^2-|\mathbf{u}_{00}^{(k)}|^{-2}|\mathbf u_{00}^{(k)*}\mathbf h_{\ell0}^{(k)}|^2\right)}{M-K+1}\nonumber\\
&=\frac{\mathbb{E}\left(|\mathbf h_{0\ell}^{(k)}|^2-|\mathbf{u}_{00}^{(k)}|^{-2}|\mathbf u_{00}^{(k)*}\mathbf h_{\ell0}^{(k)}|^2\right)}{(M-K+1)^2S^{(1)}}\approx\frac{MP_{\ell}^{(k)}\beta_{0\ell}^{(k)}\left(1-\frac{P_{\ell}^{(k)}\beta_{0\ell}^{(k)}}{S^{(k)}}\right)}{(M-K+1)^2S^{(1)}}.\nonumber
\end{align}

Then, in the case of $k>1$, for $\ell=0$, we approximate the residue intra-cell interference of ZF receivers as
\begin{align}\label{eqn:ZF_approx_3}
\mathbb{E}|\mathbf{g}_{00}^{(1)*}\mathbf h_{00}^{(k)}|^2&\approx\frac{MC^{1-\epsilon}(\lambda_\mathrm{b}\pi)^{\frac{\alpha(1-\epsilon)}{2}}\Delta^{(k)}_1}{(M-K+1)^2\left(C^{1-\epsilon}(\lambda_\mathrm{b}\pi)^{\frac{\alpha(1-\epsilon)}{2}}+\Delta^{(k)}_1\right)S^{(1)}},
\end{align}
which follows from $P_0^{(k)}\beta_{00}^{(k)}\approx\left(\mathbb{E}\left[R_{00}^{(k)2}\right]\right)^{-\frac{\alpha(1-\epsilon)}{2}}=(\lambda_\mathrm{b}\pi)^{\frac{\alpha(1-\epsilon)}{2}}$; for $\ell>0$, the out-of-cell interference is upper bounded (and approximated) as
\begin{align}\label{eqn:ZF_approx_4}
\mathbb{E}|\mathbf{g}_{00}^{(1)*}\mathbf h_{0\ell}^{(k)}|^2&\lesssim\frac{MP_1\beta^{(k)}_{0\ell}}
{(M-K+1)^2S^{(1)}}.
\end{align}
The noise term in the denominator is
\begin{align}\label{eqn:noise}
|\mathbf{g}_{00}^{(1)}|^2\sigma^2=\frac{\sigma^2}{(M-K+1)S^{(1)}}.
\end{align}
Next, substituting (\ref{eqn:approx_new1}), (\ref{eqn:approx_new2}) and (\ref{eqn:ZF_approx_1})-(\ref{eqn:noise}) for (\ref{eqn:UHF_UL_SIR}), conditioning on $R_{00}^{(1)}=x$, and after some algebraic manipulation, the SINR expression is simplified as
$$
\mathrm{SINR}=\left(C_6(\lambda_\mathrm{b}\pi x^2)^{\frac{\alpha}{2}(1-\epsilon)}+C_7(\lambda_\mathrm{b}\pi x^2)^{{\alpha}(1-\epsilon)}\right)^{-1}.
$$
The rest of the proof follows the same line as in Appendix B.
\endproof
\bibliographystyle{IEEEtran}

\end{document}